\documentclass[aps,prd,twocolumn,superscriptaddress,amsmath,amssymb,floatfix]{revtex4-2}
\usepackage[T1]{fontenc}
\usepackage{graphicx}
\usepackage{bm}
\usepackage{xcolor}
\usepackage[colorlinks=true,allcolors=blue!50!black]{hyperref}
\graphicspath{{figures/}}
\newcommand{\avS}{\langle\mathbf{S}\rangle}
\newcommand{\Ep}{\mathbf{E}_{+}}

\begin{document}

\title{Energy-flow bubbles at the Schwarzschild horizon and their Regge-pole clock}

\author{Maciej Janowicz}
\affiliation{Department of Applied Mathematics, Faculty of Informatics and Engineering,
Warsaw University of Life Sciences -- SGGW, ul.~Ciszewskiego 8, 02-786 Warsaw, Poland}
\author{Arkadiusz Orłowski}
\affiliation{Institute of Mathematics and Cryptology, Faculty of Cybernetics,
Military University of Technology, ul.~Kaliskiego 2, 00-908 Warsaw, Poland}

\date{November 15, 2026}

\begin{abstract}
For a circularly polarized plane wave scattered by a Schwarzschild black hole, the time-averaged energy flow on the
side of the horizon facing the wave contains small, sealed recirculation cells. In the main family, pairs of zero
circles are born on the axis and slide into the horizon in periodic frequency windows; at higher frequencies further
cells appear off the axis. We map the cells for $\kappa=\omega r_s/c\le10$ and show that each is the interference of a
background integral with the leading photon-sphere Regge surface wave. A cell can die only at a zero of the
circularly polarized field on the horizon, and all sixteen measured deaths are such zeros. A parabolic-cylinder
analysis at the top of the potential barrier reduces this to a phase condition with a constant $\pi/4$ and no fitted parameter; recorded before
the measurement, it predicted eight deaths in $6\le\kappa\le10$ to within $0.012$ in $\kappa$. Electromagnetic duality
makes the core of each main-family cell a circle of linear polarization; inside such a cell, flow lines lie on nested
tori and close into torus knots of one chirality.
\end{abstract}

\maketitle

\textit{Introduction.}---%
Maxwell's equations in a static gravitational field coincide with Maxwell's equations in flat space filled with an
inhomogeneous medium whose permittivity and permeability tensors are equal,
$\varepsilon^{ij}=\mu^{ij}=-\sqrt{-g}\,g^{ij}/g_{00}$~\cite{Tamm1924,Plebanski1960}. In Schwarzschild coordinates,
with horizon radius $r_s$,
\begin{equation}
  \varepsilon_r=\mu_r=1,\qquad \varepsilon_t=\mu_t=1/f,\qquad f=1-r_s/r ,
  \label{eq:medium}
\end{equation}
for the radial and tangential components (other coordinates give other, equivalent media). The medium fills the whole
exterior, with a $1/r$ tail and $\varepsilon_t\to\infty$ at $r_s$ (a feature of the Schwarzschild-coordinate
description); the horizon enters through the condition that waves there are purely ingoing. This is not a ball of
material in vacuum.

The quantity we study is the time-averaged Poynting vector $\avS=\frac12\mathrm{Re}(\mathbf E\times\mathbf H^*)$ of a
monochromatic wave in this picture. It is the Killing-energy current: its coordinate components are
$\langle S\rangle^i=J^i/(r^2\sin\theta)$, with the current density $J^i=-\sqrt{-g}\,T^i{}_t$ (physical spherical
components in the SM), and its integral curves and zeros, as unparametrized curves, coincide with those of the
energy flux measured by static observers.\footnote{See the Supplemental Material (appended below)
for the census record, proper lengths, nondegeneracy margins, the derivation of the phase condition, the knot
construction, controls, and the independent implementation.} Those observers' fields diverge at $r_s$,
while $J^i$ stays finite, so flow lines can be followed up to the horizon. Zeros, vortices, and saddles of $\avS$
organize the flow of energy around optical scatterers~\cite{Bashevoy2005,Tribelsky2022}. For black holes,
electromagnetic scattering has been studied in the far zone: cross sections, glories, and
orbiting~\cite{Mashhoon1973,Crispino2009,FHM1988}, and complex angular momentum (CAM), in which Regge poles are
surface waves on the photon sphere~\cite{AnderssonThylwe1994,Decanini2003,Decanini2010,FolacciOuldElHadj2019,DolanOttewill2009}.
Here we look at the near zone of the horizon and find small recirculation cells of the energy flow. We locate them,
follow their birth and death in frequency, and show that a Regge pole times them. The aim is an elementary account:
every feature has an ordinary cause---duality, rotational symmetry, the photon-sphere surface wave, and the parabolic top of
the potential barrier.

\textit{Model and method.}---%
We set $r_s=1$ and $\kappa=\omega r_s/c$. A plane wave of helicity $\sigma=\pm1$ travels along $+\hat{\mathbf z}$ with
time factor $e^{-i\omega t}$; ``front'' denotes the side facing the incident wave ($z<0$), and $\theta'=\pi-\theta$.
The fields separate into TE and TM Debye potentials which obey the same Regge--Wheeler equation ($s=1$). Each partial
wave is purely ingoing at $r_s$ and is normalized by its ingoing amplitude $A_\mathrm{in}e^{-i\kappa x_*}$ at
infinity, where $x_*=r+\ln(r-1)$ is the tortoise coordinate. The Coulomb-type logarithm of the long-range medium is
thus contained in $x_*$, and the remaining $l$-independent phase drops out of $\avS$. We normalize $\avS$ so that the
incident flux at infinity is $S_\mathrm{in}=1/2$. The field is summed over $l\le l_\mathrm{max}$; with
$l_\mathrm{max}=7\kappa+22$ the change on adding ten partial waves is $\le6\times10^{-7}$ in the cube $|x_i|\le5$,
and near the horizon waves with $l>b_c\kappa$ tunnel through the photon-sphere barrier, so far fewer suffice
($b_c=3\sqrt3/2$; SM).
We used Claude (Anthropic), model Claude Opus 5.5, in conversation to help formulate hypotheses, plan the numerical
experiments, and critically assess results, and through Claude Code (versions 2.1.x) to write the numerical code, the
figures, and a first draft of the text, all directed by a series of written instructions from the authors (archive
available on request). The authors reviewed the results, which were checked by internal tests and, for the field near
the cell, by an independent implementation written blind with OpenAI Codex CLI 0.147.0 (model gpt-5.6-sol; see
Verification).

\textit{Duality as a tool.}---%
In an $\varepsilon=\mu$ medium a field of helicity $\sigma$ satisfies $\mathbf H_\sigma=-i\sigma\mathbf E_\sigma$
everywhere, by uniqueness of the solution that is outgoing at infinity and ingoing at $r_s$. Hence
\begin{equation}
  \avS_\sigma=\frac{\sigma}{2}\,\mathrm{Im}\bigl(\mathbf E_\sigma^*\times\mathbf E_\sigma\bigr),
  \label{eq:spin}
\end{equation}
which is proportional to $\sigma$ times the electric spin density, and to the normal of the polarization ellipse of
$\mathbf E_\sigma$. For an incident polarization $a_+\hat{\mathbf e}_++a_-\hat{\mathbf e}_-$ the helicity cross terms
are purely imaginary, so $\avS=|a_+|^2\avS_++|a_-|^2\avS_-$, and a mirror reflection gives, in cylindrical coordinates,
\begin{equation}
  \avS=S^{+}_{\rho}\hat{\boldsymbol\rho}+S^{+}_{z}\hat{\mathbf z}+P\,S^{+}_{\varphi}\hat{\boldsymbol\varphi},
  \qquad P=|a_+|^2-|a_-|^2 ,
  \label{eq:split}
\end{equation}
with $|a_+|^2+|a_-|^2=1$. The cells below are structures of this Killing-energy current; its split into orbital and
spin parts is ambiguous in this anisotropic medium (SM). These identities are known: helicity is conserved by dual media and the energy and momentum densities split into
helicity parts~\cite{BliokhBekshaevNori2013,FernandezCorbaton2013,AlpeggianiEtAl2018}; vacuum Maxwell theory is
duality invariant in any metric~\cite{DeserTeitelboim1976}, so helicity is conserved in the scattering; and Schwarzschild scattering does not reverse the helicity
of electromagnetic waves~\cite{FHM1988}. We use them only as tools. Three consequences matter below. (i) The meridional
flow, its Stokes function $\psi$ ($\partial_\rho\psi=\rho S_z^+$, $\partial_z\psi=-\rho S_\rho^+$), and its zero
circles do not depend on polarization. (ii) For a circularly polarized wave, $\avS$ vanishes exactly on the L lines
(lines of linear polarization~\cite{Nye1983,BerryDennis2001}) of $\Ep$, not of the total field; since rotation about
$z$ multiplies $\Ep$ by a phase, its L set is axisymmetric, and zeros of $\avS$ are circles. Duality gives the
implication ``L point of $\Ep$ $\Rightarrow$ full zero of $\avS$'', not that every center of the meridional flow is an
L circle. Without duality an L circle of the field is unrelated to the zeros of $\avS$, and a center of the meridional
flow lies on one only if one extra real condition holds (codimension one); with duality every L circle is a zero,
so for an L-circle core no condition is needed. (iii) The flow is axisymmetric for every $P$. We therefore work with
$\sigma=+1$ ($P=1$) unless stated otherwise.

\begin{figure*}[t]
  \includegraphics[width=\textwidth]{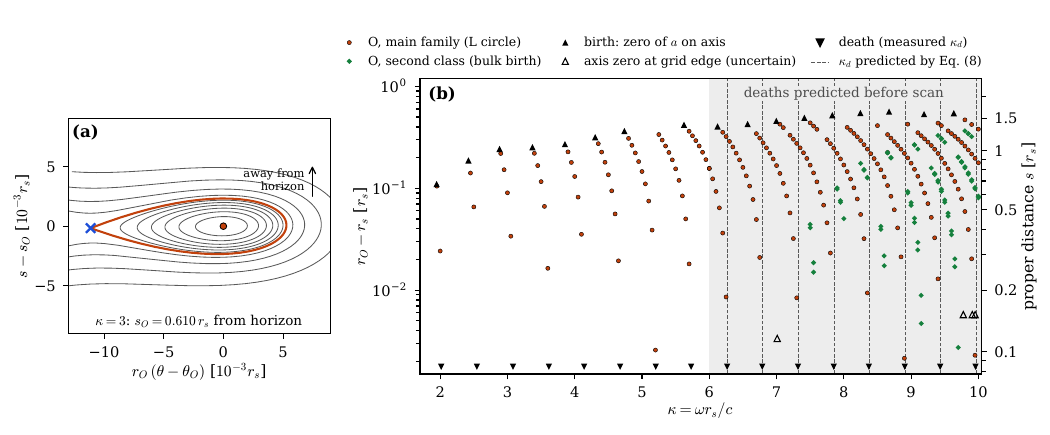}
  \caption{Recirculation cells of the time-averaged energy flow $\avS$ (circular polarization). (a)~The cell at
  $\kappa=3$ in proper lengths around its core O (equal scales): radial proper distance $s-s_O$ and tangential length
  $r_O(\theta-\theta_O)$, in $10^{-3}r_s$. Gray: level sets of the Stokes function $\psi$; red: the separatrix through
  the saddle X (cross); dot: O. The core lies at a proper distance $s_O=0.610\,r_s$ from the horizon. In
Schwarzschild coordinates the same cell is strongly elongated, $0.016\times0.0023\,r_s$.
  (b)~Coordinate distance $r_O-r_s$ of the O circles from the horizon versus $\kappa$ (right axis: proper distance
  $s$). Red: main family (O is an L circle of $\Ep$); green: second class; $\blacktriangle$: zeros of $\Ep$ on the axis
  (births; $\triangle$: zeros at the edge of the grid, uncertain); $\blacktriangledown$: measured deaths $\kappa_d$
  (sixteen, $m=4$--$19$); dashed lines: deaths predicted by Eq.~\eqref{eq:death}, recorded before the scan of the
  shaded range. Figure
  generated by AI-written scripts checked by the authors (see Model and method).}
  \label{fig:census}
\end{figure*}

\textit{Census of zero circles.}---%
Table~\ref{tab:census} lists where we searched for O (center) and X (saddle) circles of the meridional flow, with which
truncation, and what we found. There are none outside a zone near the front pole, $1<r<1.8$, $\theta'<1$: none behind
the hole up to $z=50$ (where the wave is focused by the $1/r$ tail) and none in front from $z=-20$ to $-5$, out to
$\rho=10$, at $\kappa=3$ (rear also at $\kappa=2.5$).
Near the front pole, up to $\kappa\approx5.6$ the flow has at most one O--X pair at a time, which exists only in windows
of $\kappa$; at $\kappa\approx5.65$--$5.70$ the next pair is born before the previous one dies.
For larger $\kappa$ the windows overlap, and up to three pairs of this main family coexist. From $\kappa\approx7.5$ a
second class appears: an O--X pair, both circles off the axis and not L circles, is born in the bulk and turns into a
quartet of two counter-rotating O circles and two L saddles, which ends in the bulk or exits through the horizon. At
$\kappa\approx9.4$--$9.9$ six to seven O--X pairs coexist (Fig.~\ref{fig:census}b).
We resolved one second-class window to $0.01$ in $\kappa$ and the others only to $0.05$; we treat this class as a
preliminary observation (SM) and concentrate here on the main family.

\begin{table}[tb]
\caption{Census of O and X circles of the meridional flow ($P$-independent). $\theta'=\pi-\theta$; ``main'': O is an
L circle of $\Ep$, X is not; ``2nd class'': bulk-born pairs and quartets of two counter-rotating O and two L saddles (SM). Details and grids in the SM.}
\label{tab:census}
\begin{ruledtabular}
\footnotesize
\begin{tabular}{@{}llll@{}}
region & $\kappa$ (step) & $l_\mathrm{max}$ & found\\
\hline
$|x_i|\le5$ & $0.5$--$6$ ($0.25$) & $7\kappa{+}22$ & front pole only\\
$1<r<1.8$, $\theta'<1$ & $1.7$--$6$ ($0.05$) & $2\kappa{+}16$ & 1 main pair$^{a}$\\
$1<r<1.8$, $\theta'<1$ & $6$--$10$ ($0.05$) & $\lfloor2.5\kappa\rfloor{+}20$ & 1--3 main, 2nd cl.\\
$z\in[0,50]$, $\rho\le10$ & $3$, $2.5$ & 185--200$^{b}$ & none\\
$z\in[-20,-5]$, $\rho\le10$ & $3$ & 185--200$^{b}$ & none\\
\end{tabular}
\end{ruledtabular}
\raggedright\footnotesize $^{a}$Two at $\kappa\approx5.65$--$5.70$.
$^{b}$Two truncations each ($200/185$ at $\kappa=3$, $170/155$ at $\kappa=2.5$), identical results.
\end{table}

\textit{Birth and death.}---%
On the axis the helicity-$+$ field is $\Ep=a(z)\hat{\mathbf e}_+$, circularly polarized or zero. An L circle can
therefore shrink to the axis or emerge from it only at an isolated zero of $a(z,\kappa)$: a zero of the field on the
axis, which is otherwise a C line, where the backscattered wave exactly cancels the incident one. Each pair of the main
family is born there: the core (an L circle) emerges together with the saddle. Births lie at $\kappa^*=1.947$, $2.420$,
$2.883$, \dots; their spacing decreases from $0.49$ to $0.42$ over the range studied, while deaths come every
$\approx0.53$, so the lifetime of a pair grows, from $0.07$ for the first pair to $1.28$ for the pair that dies at
$\kappa=9.96$.

At $\kappa=3$ the core lies at a proper distance $0.610\,r_s$ from the horizon ($r_O-r_s=0.090$), and the cell is a
flat lens of $4.7\times10^{-3}\times1.6\times10^{-2}\,r_s$ (radial $\times$ tangential, proper); for $\kappa=2$--$6$
the cores lie $0.31$--$0.81\,r_s$ from the horizon (SM).
The circulating Killing-energy flux, $2\pi|\psi_O-\psi_X|$, is $1.0\times10^{-6}$ of the flux through $\pi r_s^2$ at
$\kappa=3$.

A pair dies by sliding into the horizon: its coordinate size shrinks in proportion to $r-r_s$, so the bubble keeps its
shape in the tortoise coordinate (O--X distance $/(r_O-r_s)=0.18$ at $\kappa\approx2$, $0.07$--$0.10$ for
$\kappa=6$--$10$). O and X do not collide; the cell leaves through the boundary.

\begin{figure}[tb]
  \includegraphics[width=\columnwidth]{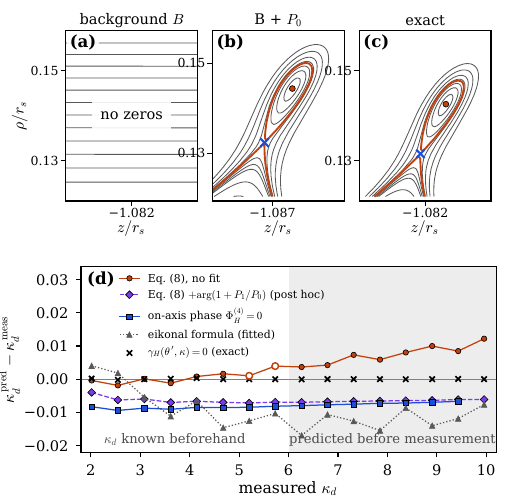}
  \caption{(a)--(c) Level sets of the Stokes function $\psi$ of $\avS$ at $\kappa=3$ near the cell, computed from the
  background integral $B$ alone (no zeros), from $B+P_0$ (own window; shifted by $5.4\times10^{-3}r_s$), and from the
  exact partial-wave sum. Coordinate axes, not to scale (the $\rho$ range is $5.4$ times the $z$ range).
  (d)~Predicted minus measured death frequencies for the $16$ measured deaths of the main family ($m=4$--$19$).
  Circles: Eq.~\eqref{eq:death}, no fitted parameter (open: $m=10$, $11$, measured after the prediction test);
  diamonds: Eq.~\eqref{eq:death} plus the $n=1$ pole in WKB form, added post hoc; squares: on-axis phase
  $\Phi_H^{(4)}=0\bmod2\pi$; triangles: eikonal formula of the earlier draft (fitted constant); crosses: zeros of
  $\gamma_H$, Eq.~\eqref{eq:exact}. Shaded: deaths predicted before they were measured.
  Figure generated by AI-written scripts checked by the authors (see Model and method).}
  \label{fig:cam}
\end{figure}

\textit{Regge poles and the phase condition.}---%
On the front axis the partial-wave sum for $a(r)$ is non-alternating: the factor $(-1)^l$ of $P_l(-1)$ cancels against
$i^{2l}$ from the incident-wave coefficients and the ingoing normalization. Writing $a(r)=\sum_{l\ge1}h(l+\tfrac12)$, the
Watson--Sommerfeld kernel is therefore $\tan\pi\lambda$, and deforming the contour as in Mie
theory~\cite{Nussenzveig1992} gives
\begin{equation}
  a(r)=B(r)+\sum_{n\ge0}P_n(r),\qquad
  P_n=-\pi i\,\mathrm{Res}_{\lambda_n}h\;K(\lambda_n),
  \label{eq:cam}
\end{equation}
with $K(\lambda)=e^{i\pi\lambda}/\cos\pi\lambda$. Here $\lambda_n(\kappa)\approx b_c\kappa+i(n+\frac12)$ are the Regge
poles (zeros of $A_\mathrm{in}$, from Leaver's continued fraction~\cite{Leaver1985}) and $B$ is the background integral
over continuous $\lambda$ (real and imaginary axes). The decomposition assumes no contribution from arcs at infinity
and no poles in the fourth quadrant; both were verified numerically, not proven. With five poles it reproduces the
partial-wave sum on the front axis to $7.5\times10^{-5}$ (maximum; median $2.6\times10^{-8}$) of the incident
amplitude for $\kappa\in[1.5,5]$; off-axis formulas and accuracy up to $\kappa=10$ are in the SM.
The background has $|B|-1\in[0,0.010]$, while near the horizon the leading surface wave is stronger,
$|P_0|=1.1$--$2.0$ at $r=1.02$.
Off the axis the sums remain non-alternating near the front pole, and $B$ alone or the poles alone produce no zero of
$\avS$ there. $B+P_0$ produces the bubble, shifted by $5.4\times10^{-3}r_s$, about a third of its coordinate length
($1.6\times10^{-2}r_s$), and $B+P_0+P_1$ places it to $3.2\times10^{-4}r_s$ (Fig.~\ref{fig:cam}a--c). The numerical
Watson--Sommerfeld decomposition thus shows that the bubble is the interference of the background with the
photon-sphere surface wave.

\textit{Death of a cell.}---%
Near the horizon every partial wave behaves as $t^{-i\kappa}[1+O(t)]$, $t=r-r_s$. The tangential field there is
therefore circularly polarized,
$\mathbf E_t\simeq t^{-i\kappa}\gamma_H(\theta')(\hat{\boldsymbol\theta}-i\hat{\boldsymbol\varphi})/\sqrt2$, with an
amplitude $\gamma_H(\theta')$ that depends only on $A_\mathrm{in}(\lambda)$ (SM), and $S_r=-|\gamma_H|^2/2+O(t)$.
A zero of the meridional flow can reach the horizon only where $\gamma_H$ vanishes, so a cell can die only at
\begin{equation}
  \gamma_H(\theta'_d;\kappa_d)=0 ,
  \label{eq:exact}
\end{equation}
one complex equation for two real unknowns. All sixteen measured deaths of the main family are such zeros, to within
their uncertainty ($\le3\times10^{-4}$), at $\theta'_d=0.20$ to $0.07$; five further pairs of zeros are
the exits of second-class quartets through the horizon (SM).

Applied to $\gamma_H(\theta')$, the Watson--Sommerfeld transformation gives $\gamma_H=B_H+\sum_nP_{n,H}$, with the same
poles and angular factors $P_{\lambda_n-1/2}(\cos\theta')$. On the axis define the phase
$\Phi_H^{(N)}=\arg\sum_{n<N}P_{n,H}-\arg(-B_H)$, in particular $\Phi_H^{(1)}=\arg P_{0,H}-\arg(-B_H)$. Its exact
decomposition is
$\Phi_H^{(1)}=\pi/2+\arg\lambda_0+\arg K(\lambda_0)-\arg A_\mathrm{in}'(\lambda_0)-\arg I$, where $B_H=I/(i\kappa)$.
The phases of $A_\mathrm{in}'$ and $I$ follow from the solution at the top of the
barrier, $r=3/2$. There the Regge--Wheeler potential is parabolic in $x_*$, the ingoing solution is a
parabolic-cylinder function, and matching it to WKB solutions on both sides gives
\begin{equation}
  A_\mathrm{in}(\lambda)\simeq e^{i\kappa[1+D(r_s)]}\,e^{ia(2\ln c+J)}\,e^{\pi a/2}\,\frac{\sqrt{2\pi}}{\Gamma(\frac12+ia)},
  \label{eq:ain}
\end{equation}
where $a\propto\lambda^2-\frac14-b_c^2\kappa^2$ measures the distance from the barrier top in units of its curvature,
$c=(64L/729)^{1/4}$ is the barrier scale, and $J=6\ln3-3\ln(2+\sqrt3)$ (SM). The poles of $\Gamma$, $a=i(n+\frac12)$,
are the Regge poles; $|A_\mathrm{in}|^2=1+e^{2\pi a}$ is the parabolic-barrier transmission. The phase $D$ is
\begin{equation}
  D(r)=\int_r^\infty\Bigl[1-\Bigl(1-\frac{b_c^2f}{x^2}\Bigr)^{1/2}\Bigr]\frac{dx}{f},
  \label{eq:D}
\end{equation}
with $D(r_s)=3\sqrt3-\frac92+6\ln\frac{2+\sqrt3}{2}=4.4390$, the WKB phase deficit, per unit $\kappa$, of the wave
with the critical angular momentum $b_c\kappa$, which passes over the barrier top; it is not the path of a ray (SM).
At $a=i/2$, Eq.~\eqref{eq:ain} gives $\arg A_\mathrm{in}'(\lambda_0)=\kappa(1+D)+3\pi/4$, and for the background
$\arg I\to-\kappa-\pi/2$. Setting $\Phi_H^{(1)}=2\pi m$ then gives
\begin{equation}
  2\pi\,\mathrm{Re}\,\lambda_0-\arg\!\bigl(1+e^{2\pi i\lambda_0}\bigr)+\arg\lambda_0-\kappa D(r_s)+\frac{\pi}{4}=2\pi m .
  \label{eq:death}
\end{equation}
The constant $\pi/4$ comes from the factor $e^{\pi a/2}$. Numerically, the smooth part of $\Phi_H^{(1)}-[2\pi\mathrm{Re}\lambda_0-\kappa D(r_s)]$ tends to $0.785399$ as $\kappa\to\infty$,
against $\pi/4=0.785398$; the oscillating term $-\arg(1+e^{2\pi i\lambda_0})$, of amplitude $|e^{2\pi i\lambda_0}|\approx e^{-\pi}=0.043$~rad, is known
exactly from $\lambda_0$. Equation~\eqref{eq:death} contains no fitted parameter; only the pole $\lambda_0(\kappa)$
(real and imaginary parts) is computed numerically. A remaining $c_1/\kappa$ term in the phase ($c_1\approx0.10$) is
known only from the numerics and shifts $\kappa_d$ by about $-0.001$.

Equation~\eqref{eq:death} is thus the leading approximation to Eq.~\eqref{eq:exact}: one pole, on the axis, leading
WKB order. Its two main errors are known. (i) The $n=1$ pole: at the horizon
$P_{1,H}/P_{0,H}\simeq i\,(4\sqrt3/9)\,e^{J-2\pi}\kappa=0.020\,i\kappa$ (WKB, within $1\%$ of the exact ratio), which
shifts the phase by $\approx0.02\kappa$~rad, i.e., $\kappa_d$ by $+0.0017\kappa$. (ii) The angle: the cell dies off
the axis, where the surface wave has decayed to the background level, and there the phases of $P_{0,H}$ and $B_H$
relative to the axis differ by $0.09$--$0.10$~rad, which shifts $\kappa_d$ by $-0.007$ to $-0.008$. Below
$\kappa\approx5$ the two errors largely cancel.

\textit{Prediction test.}---%
Before scanning $\kappa\in[6,10]$ we recorded the deaths predicted by Eq.~\eqref{eq:death}, $m=12$--$19$. The scan
found eight deaths, in the predicted order. All eight lie within $+0.004$ to $+0.012$ of the predictions
(mean $+0.0075$; Fig.~\ref{fig:cam}d), and the irregularity of their spacings, $\pm0.006$, is reproduced; it comes from
the oscillating term, which the eikonal spacing does not have. Our earlier scan of $\kappa\le6$ had not resolved the
deaths $m=10$, $11$ of two overlapping windows, and our earlier table labeled the predicted deaths $m=10$--$17$; we
measured these two deaths after the test, $\kappa_d=5.2036$ and $5.7300$, and Eq.~\eqref{eq:death} misses them by
$+0.0010$ and $+0.0039$.
The error of Eq.~\eqref{eq:death} grows by $0.0016$ per unit $\kappa$, the $n=1$ term (i); adding it in WKB form
leaves a constant $-0.0064\pm0.0007$, the angular term (ii) (Fig.~\ref{fig:cam}d). The on-axis criterion
$\Phi_H^{(N)}=0$ converges for $N\ge3$ to an offset of $-0.008$ to $-0.007$, which survives extrapolation to $r\to r_s$
(SM); the off-axis condition $B_H+\sum_{n<N}P_{n,H}=0$ with $N\ge3$ reproduces the deaths to $3\times10^{-4}$. The mean
spacing of the sixteen deaths is $0.52944\pm0.00002$, against $0.53028$ from Eq.~\eqref{eq:death}.

Births need, on the axis at finite $r$, both $\Phi=\arg P_0-\arg(-B)=2\pi m$ and $|P_0/B|=1$. The same WKB construction gives the modulus in
closed form, $|P_0|\approx0.925\sqrt\kappa$ at the horizon (exact $1.595$ at $\kappa=3$).
Since $|B_H|\approx1$, $|P_0/B|$ grows with $\kappa$, which is why the windows widen. This model reproduces the seven
on-axis births for $\kappa\le4.74$ to $\Delta\kappa\le0.014$; it is a postdiction, and we did not apply it to
$\kappa>6$ before the scan.

Regge poles have been used for radiation generated by sources in the strong-field region~\cite{FolacciOuldElHadj2018,FolacciOuldElHadj2020},
but those works describe waveforms at infinity. To our knowledge, Regge poles have not been used to describe the
near-zone field structure.

\textit{Tori and knots.}---%
We assume a single plane wave, the Schwarzschild metric, a lossless exterior, and the time-averaged flow. Then
$\nabla\cdot\avS=0$ and the axisymmetry of Eq.~\eqref{eq:split} reduce the flow to a one-degree-of-freedom Hamiltonian
system in $(\rho,z)$ with Hamiltonian $\psi$, plus a quadrature for $\varphi$~\cite{MezicWiggins1994,HallerMezic1998}.
Inside a cell, flow lines lie on nested tori $\psi=\mathrm{const}$; elsewhere the level sets are open tubes. For
$P\ne0$ the X circle is a periodic orbit and the separatrix is a surface of revolution, which would split under
generic symmetry-breaking perturbations. Let $q(\psi)=2\pi/\Delta\varphi(\psi)$ be the number of poloidal turns per
toroidal turn, with $q(\psi;P)=q(\psi;1)/P$. For a cell of the main family three statements give knots.
(a) The core is an L circle of $\Ep$ (all $51$ frequencies of the dense scan with a cell), with winding number
$|\deg\mathbf b|=1$ of the minor-axis field $\mathbf b$ of the polarization ellipse (11 cells); so $S^+_\varphi$
vanishes linearly at O and $q\to\infty$.
(b) X is hyperbolic, its separatrix is a homoclinic loop, and $S^+_\varphi(\mathrm X)\ne0$; the transit time past X
then diverges logarithmically and $q\to0$ as $q\simeq2\pi/(a'+b'\ln\delta^{-1})$, with
$\delta=|\psi-\psi_X|/|\psi_O-\psi_X|$ and $b'=S^+_\varphi/(\rho\lambda_X)$ at X (checked to $10^{-4}$, SM).
(c) By continuity, $q$ takes every value in $(0,\infty)$ (Darboux). On a torus with $q=m/n$, $\gcd(m,n)=1$, every
flow line closes after $m$ poloidal and $n$ toroidal turns into a torus knot $T(m,n)$. Since $\Delta\varphi$ has a
fixed sign in every cell we examined, the knots have one chirality for a given sign of $P$, and reversing the
helicity ($P\to-P$) reverses it. Since $T(m,n)\cong T(n,m)$, the levels $q>1$ (near the core) already give every
torus-knot type; the thin layer $q<1$ at the separatrix repeats them.
The mechanism is classical for swirling vortex rings~\cite{Moffatt1969,EncisoPeraltaSalas2012,KediaEtAl2016}. What is
specific here is that duality makes the L core a full zero of $\avS$ for every $P$, so the divergence of $q$ at the
core is structurally stable; without duality an L core is a coincidence of codimension one. These knots are
stationary streamlines of $\avS$, not the instantaneous field lines or phase singularities of knotted light
fields~\cite{IrvineBouwmeester2008,Dennis2010}.

\begin{figure}[tb]
  \includegraphics[width=\columnwidth]{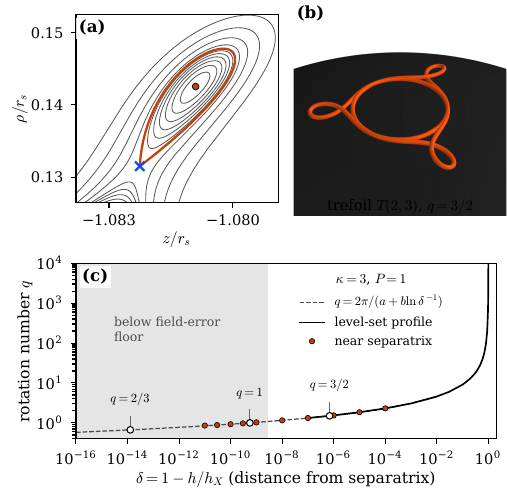}
  \caption{(a) Cell at $\kappa=3$: level sets of $\psi$ (meridional sections of the tori of $\avS$), core O (an L
  circle of $\Ep$), saddle X, and the homoclinic separatrix (red). Coordinate axes, not to scale ($\rho$ range $5.3$
  times the $z$ range). (b)~Closed flow line of $\avS$ with $q=3/2$, a trefoil $T(2,3)$, drawn above the horizon
  (black); the meridional cross section is enlarged by the dilation $(\rho,z)\mapsto\mathrm O+a(\rho-\rho_O,z-z_O)$,
  an isotopy away from the axis that does not change the knot type. (c)~Rotation number $q$ versus $\delta=1-h/h_X=|\psi-\psi_X|/|\psi_O-\psi_X|$ ($P=1$): level-set profile,
  separate points near the separatrix, and the logarithmic law. Shaded: below the field-error floor. Figure generated
  by AI-written scripts checked by the authors (see Model and method).}
  \label{fig:tori}
\end{figure}

Hypotheses (a) and (b) have strong numerical support: the local nondegeneracy margins exceed the estimated field
error by at least $2.5\times10^5$ in all $14$ main-family cells tabulated ($\kappa=2$--$9$). These are floating-point
estimates; they do not cover the global structure of the separatrix, which rests on the census (SM). Since every knot
type occurs at $q>1$, the knot statement does not depend on the layer $q<1$.
The field error does limit the construction of particular knots: levels $\delta\lesssim3\times10^{-9}$ cannot be
distinguished from it. At $\kappa=3$, $q=1$ lies at $\delta=5.6\times10^{-10}$, below this floor, while $T(2,3)$ at
$q=3/2$ lies at $\delta\approx7\times10^{-7}$, above it (Fig.~\ref{fig:tori}c). We built six knots from $T(2,3)$ to
$T(5,8)$ at $\kappa=3$, each closing in full 3D integration to $\le2.4\times10^{-4}$ of the torus radius (SM). Even
$T(2,3)$ lies within $10^{-6}$ of the separatrix in $\psi$, the part of the cell most fragile to symmetry breaking.

\textit{Verification.}---%
The field near the cell was recomputed by a second implementation, written by a different AI system (OpenAI Codex)
under a blind protocol, with one documented deviation (a file listing of path names before the rules were set; SM):
it read only the task specifications, derived the radial problem itself (Frobenius series at
the horizon, DOP853 integration, Jost-function matching at large $r$), froze its results by SHA-256 hashes, and only
then opened our data. At $\kappa=3$ the positions of O and X agree to $8\times10^{-11}r_s$, $\avS$ on a grid around
the cell to a median relative difference of $8\times10^{-9}$, and the absorption cross section to $1.2\times10^{-9}$.
Our own field satisfies Maxwell's equations in the medium with residuals falling as $h^4$ with the
finite-difference step (SM).
The knot construction, the CAM analysis, and the death condition were tested only internally.

\textit{Discussion.}---%
The near-horizon energy flow has striking features---sealed cells, knotted streamlines, frequency windows---but each
has an ordinary origin. Duality fixes the polarization-independent skeleton and makes the cell cores L circles;
rotational symmetry gives tori; the photon-sphere surface wave produces the cells by interference with the background;
and the parabolic top of the barrier sets the phase constant $\pi/4$ of the approximate death condition, while the
exact condition requires a zero of the circularly polarized field on the horizon. These statements concern the main family; the
off-axis cells that appear for $\kappa\gtrsim7.5$ are a preliminary observation (SM). The structure is tied to the
energy current: the canonical (orbital) momentum density has no O--X pair at the cell, but its definition in this
medium is ambiguous (SM).

What does the black hole add to dual spheres in nano-optics? The skeleton of the flow is almost empty, one cell per
window up to $\kappa\approx5.6$, against dozens for a Buchdahl star of radius $1.2\,r_s$ (SM). With waves purely
ingoing at the horizon, what remains near it is the interference of the background with the photon-sphere surface
wave, whose Regge pole fixes the phase of a structure a fraction of $r_s$ from the horizon. The poles that describe
glories and orbiting in the far zone thus also time the near-zone flow.
Our results hold for a single plane wave, a nonrotating hole, a lossless exterior, and $\kappa\le10$. For Kerr the
equivalent medium is bianisotropic; it stays dual when $\varepsilon=\mu$ and the two magnetoelectric terms are equal,
which remains to be checked, and then the analysis should carry over for incidence along the axis. Near the horizon the ratio of successive Regge-pole
terms grows as $\kappa$, so the one-pole Eq.~\eqref{eq:death} loses its meaning at $\kappa\sim50$, where
Eq.~\eqref{eq:exact} needs many poles; the limit $\kappa\to\infty$ remains open.

\begin{acknowledgments}
MJ would like to dedicate this work to the memory of his Mother and also to the community of Brazilian physicists of
Rio de Janeiro as represented by Professor Luiz Davidovich and the late Professor Herch Moys\'es Nussenzveig.
We thank Claude (Anthropic) for substantial help with this work: in conversation, Claude helped to formulate
hypotheses, design the numerical experiments, and critically review intermediate results; through Claude Code it
wrote and ran the partial-wave, complex-angular-momentum, and flow-topology codes, produced the figures, and drafted
the manuscript. We also thank OpenAI Codex for the independent implementation described in the Supplemental
Material. The authors reviewed every result reported here and take full responsibility for the content.
\end{acknowledgments}

\bibliography{refs}

\clearpage
\onecolumngrid
\begin{center}
\textbf{\large Supplemental Material}
\end{center}
\bigskip
\setcounter{equation}{0}
\setcounter{figure}{0}
\setcounter{table}{0}
\setcounter{section}{0}
\renewcommand{\theequation}{S\arabic{equation}}
\renewcommand{\thefigure}{S\arabic{figure}}
\renewcommand{\thetable}{S\arabic{table}}
\renewcommand{\theHequation}{S\arabic{equation}}
\renewcommand{\theHfigure}{S\arabic{figure}}
\renewcommand{\theHtable}{S\arabic{table}}
\renewcommand{\theHsection}{S\arabic{section}}
Conventions: $r_s=1$, $x=r/r_s$, $\kappa=\omega r_s/c$, time factor $e^{-i\omega t}$, incident wave
$\mathbf E=\hat{\mathbf e}_\sigma e^{i\kappa z}$ at infinity, $\avS=\frac12\mathrm{Re}(\mathbf E\times\mathbf H^*)$,
$S_\mathrm{in}=1/2$, $f=1-1/x$, tortoise coordinate $x_*=x+\ln(x-1)$, $\lambda=l+\frac12$, $L=l(l+1)$,
$b_c=3\sqrt3/2$, $\theta'=\pi-\theta$. ``Front'' is the side facing the incident wave ($z<0$). All lines, cells, tori
and knots are those of the time-averaged $\avS$, i.e., of the Killing-energy current, not of the instantaneous
$\mathbf E$ or $\mathbf H$. For dielectric control spheres of radius $a$, $\kappa=\omega a/c$.
Membrane regularization: $\bar f=1-1/(x-i\epsilon)$, $\epsilon>0$.

\textit{Use of AI tools.} Claude (Anthropic), model Claude Opus 5.5 (\texttt{claude-opus-5-5}), was used in
conversation to help formulate hypotheses, plan the numerical experiments, and critically assess results, and through
Claude Code (versions 2.1.x, same model) to write the partial-wave, CAM, and flow-topology codes, all figures, and a
first draft of the text. The work was directed by a series of written instructions from the authors (archive available
on request), and the authors reviewed all results. The independent implementation of Sec.~\ref{sec:codex} was written
with OpenAI Codex CLI 0.147.0, model gpt-5.6-sol, under a blind protocol.

\section{\texorpdfstring{Partial waves, the incident wave, and convergence in $l_\mathrm{max}$}{Partial waves, incident wave, convergence}}
\label{sec:pw}
For the medium $\varepsilon_r=\mu_r=1$, $\varepsilon_t=\mu_t=1/f$ the field splits into TE and TM Debye potentials.
Both radial functions obey $\frac{d}{dx}(fu')+(\kappa^2/f-L/x^2)u=0$, which after multiplication by $f$ is the
Regge--Wheeler equation $u_{**}+(\kappa^2-fL/x^2)u=0$ for $s=1$. At the horizon we take the purely ingoing solution
$u=t^{-i\kappa}\sum_nb_nt^n$, $t=x-1$, and match it to $A_\mathrm{in}e^{-i\kappa x_*}+A_\mathrm{out}e^{i\kappa x_*}$.

\textit{Incident wave.} The medium has a $1/r$ tail, so the incident wave is Coulomb-distorted. Each partial wave is
normalized by its ingoing amplitude $A_\mathrm{in}$ in the tortoise coordinate and given the flat-space coefficient
($a_l=-i^l(2l+1)/(\kappa L)$ for TM). The logarithmic phase of the long-range field is contained in $x_*$; what is left
is an $l$-independent phase, which cancels in $\avS$, so the topology and positions of zeros do not depend on it.
$S_\mathrm{in}=1/2$ is the normalization at infinity. The ``vacuum check'' $\avS=\hat{\mathbf z}/2$ (to
$8\times10^{-15}$) is a flat-space test of the code; the normalization in Schwarzschild is checked by the absorption
cross section, Sec.~\ref{sec:checks}, and by the far-field values in Sec.~\ref{sec:census}.

\textit{Comparison objects} (used only as controls, Sec.~\ref{sec:controls}): a PEC wall at $x=1+10^{-3}$; a
constant-density star with $R=1.2$, whose interior medium is $\varepsilon_r=\mu_r=1/(A\sqrt B)$,
$\varepsilon_t=\mu_t=\sqrt B/A$ with $A=\sqrt{-g_{00}}$ and $B=g_{rr}$ (the stellar matter is assumed transparent and
coupled to light only through the metric); a dielectric sphere $n=4$; and a dual sphere $\varepsilon=\mu=2$.

\textit{Convergence.} $l_\mathrm{max}=7\kappa+12$ gives an error of $7.5\times10^{-3}$ at the corners of the cube
$|x_i|\le5$; $l_\mathrm{max}=7\kappa+22$ gives $\le6\times10^{-7}$ (comparison with $l_\mathrm{max}+10$), for all
objects. Radial splines: $\le2\times10^{-9}$. Near the horizon, partial waves with $l>b_c\kappa-\frac12$ tunnel through
the photon-sphere barrier: at the core O ($\kappa=3$) the contribution of partial wave $l$ to $|\Ep|$ falls from $0.39$
($l=8$) to $9.0\times10^{-4}$ ($l=10$), $9.6\times10^{-11}$ ($l=15$) and $2.9\times10^{-16}$ ($l=20$). Near-field models:
$l_\mathrm{max}=2\kappa+16$, $r_\mathrm{max}=3$ for $\kappa\le6$ (same O, X and $\psi$ as $l_\mathrm{max}=43$ at
$\kappa=3$, difference $0$ in floating-point arithmetic) and $\lfloor2.5\kappa\rfloor+20$ for $6\le\kappa\le10$
(identical to $4\kappa+30$ at $\kappa=6,8,10$; the tail $l>l_\mathrm{max}$ is $\sim10^{-20}$ of the leading terms).
Convergence of the cell itself is shown in Table~\ref{tab:conv}.

\section{\texorpdfstring{Duality: helicity decomposition and $\avS_\sigma=(\sigma/2)\,\mathrm{Im}(\mathbf E^*\times\mathbf E)$}{Duality: helicity decomposition}}
\label{sec:duality}
These statements are short consequences of dual electromagnetism~\cite{BliokhBekshaevNori2013,FernandezCorbaton2013,AlpeggianiEtAl2018}
and of the absence of helicity reversal in Schwarzschild scattering for $s=1$~\cite{FHM1988}; we list them because
the Letter uses them.

\textit{Lemma 1.} Let $\varepsilon=\mu$ (tensors, possibly complex). Maxwell's equations
$\nabla\times\mathbf E=ik\mu\mathbf H$, $\nabla\times\mathbf H=-ik\varepsilon\mathbf E$ are then invariant under the
duality rotation $(\mathbf E,\mathbf H)\to(\mathbf H,-\mathbf E)$, so every solution splits into helicity parts
$\mathbf F_\pm=\frac12(\mathbf E\pm i\mathbf H)$, each of which satisfies $\nabla\times\mathbf F_\pm=\pm k\mu\mathbf F_\pm$;
a solution with $\mathbf F_-=0$ (helicity $\sigma=+1$) has $\mathbf H=-i\mathbf E$, and similarly for $\sigma=-1$.
An incident wave of helicity $\sigma$ satisfies $\mathbf H=-i\sigma\mathbf E$. The scattering solution is unique once it is required to be outgoing (apart
from the incident wave) at infinity and purely ingoing at $r=r_s$; each partial wave is then fixed by the Regge--Wheeler
equation, and the TE/TM pair of a helicity eigenstate is fixed with it. Hence $\mathbf H_\sigma=-i\sigma\mathbf E_\sigma$
everywhere, and
\begin{equation}
\avS_\sigma=\tfrac12\mathrm{Re}(\mathbf E_\sigma\times\mathbf H_\sigma^*)=\tfrac{\sigma}{2}\mathrm{Im}(\mathbf E_\sigma^*\times\mathbf E_\sigma).
\end{equation}
With $\mathbf E_\sigma=e^{i\alpha}(\mathbf A+i\mathbf B)$, $\mathrm{Im}(\mathbf E^*\times\mathbf E)=2\mathbf A\times\mathbf B$,
which is normal to the polarization ellipse and proportional to its area. Since $\mathbf H^*\times\mathbf H=\mathbf E^*\times\mathbf E$,
$\avS_\sigma$ is proportional to $\sigma$ times the electric spin density.

\textit{Lemma 2 (no helicity interference).} For $\mathbf E=a_+\mathbf E_++a_-\mathbf E_-$ the cross term in
$\mathbf E\times\mathbf H^*$ is purely imaginary, so $\avS=|a_+|^2\avS_++|a_-|^2\avS_-$; the relative phase of $a_\pm$
(orientation of the polarization ellipse) drops out.

\textit{Lemma 3 (poloidal--toroidal split).} Rotation by $\alpha$ about $z$ multiplies $\mathbf E_\pm$ by $e^{\pm i\alpha}$,
so $\avS_\pm$ is axially symmetric, and so is the set of L points of $\Ep$ (a phase does not change L-ness). The
reflection $\sigma_y:y\to-y$ maps $\hat{\mathbf e}_+$ onto $\hat{\mathbf e}_-$ with
$\mathbf H'=-\sigma_y\mathbf H(\sigma_y\mathbf x)$; since $(\sigma_y\mathbf a)\times(\sigma_y\mathbf b)=-\sigma_y(\mathbf a\times\mathbf b)$,
$\avS_-(\mathbf x)=\sigma_y\avS_+(\sigma_y\mathbf x)$, i.e., $S_\rho^-=S_\rho^+$, $S_z^-=S_z^+$, $S_\varphi^-=-S_\varphi^+$. Hence
$\avS(P)=S_\rho^+\hat{\boldsymbol\rho}+S_z^+\hat{\mathbf z}+PS_\varphi^+\hat{\boldsymbol\varphi}$.

\textit{Zeros.} For $P\ne0$ a full zero of $\avS$ requires $\avS_+=0$, i.e., an L point of $\Ep$, not of the total
field; for $0<|P|<1$ the X circle is not a zero, since $S_\varphi=PS_\varphi^+(\mathrm X)\ne0$. For any dual scatterer,
axisymmetric or not, the zeros of $\avS$ of a pure-helicity field are the L lines of $\mathbf E$, which are generic
curves in 3D; this contrasts with the isolated zeros of non-dual particles~\cite{Bashevoy2005,Tribelsky2022}.

\textit{Lossy case.} In $\bar f$ the decomposition remains exact but $\nabla\cdot\avS=-(k/2)\,\mathrm{Im}(1/\bar f)(|\mathbf E_t|^2+|\mathbf H_t|^2)<0$;
$\psi$ is no longer conserved, and for $P\ne0$ the core becomes an attracting or repelling invariant circle.

\textit{Numerical checks.} $\max|\mathbf H_++i\mathbf E_+|/|\mathbf E_+|=0$ in floating-point arithmetic (horizon,
$\bar f$, Buchdahl, dual sphere), against a median of $0.51$ over the window for the PEC sphere.
$\max|S_\varphi|/|\avS|\le4.5\times10^{-16}$ at linear polarization; $S_\mathrm{pol}(P)=S_\mathrm{pol}^+$ and
$S_\varphi(P)=PS_\varphi^+$ to $\le5\times10^{-15}$ for $P\in\{0,0.25,0.5,1\}$ and three relative phases.

\section{Census of zero circles: domain, families, and the second class}
\label{sec:census}
Table~\ref{SM-tab:census} gives the full search record.

\begin{table}[ht]
\caption{Where O/X circles of the meridional flow were searched for, with which truncation, and what was found.}
\label{SM-tab:census}
\begin{ruledtabular}
\begin{tabular}{p{4.0cm}p{2.4cm}p{3.3cm}p{2.9cm}p{4.1cm}}
region & $\kappa$ & model, $l_\mathrm{max}$ & grid & found\\
\hline
$|x_i|\le5$ & $0.5$--$6$ (step $0.25$) & full, $7\kappa+22$ & $3$D search & no O/X circles except near the front pole\\
front, $1<r<1.8$, $\theta'<1.0$ & $1.70$--$6.00$ (step $0.05$) & near field, $2\kappa+16$, $r_\mathrm{max}=3$ & polar, Newton & one main pair (two at $5.65$--$5.70$)\\
front, $1<r<1.8$, $\theta'<1.0$ & $6.00$--$10.00$ (81 values) & near field, $\lfloor2.5\kappa\rfloor+20$ & polar $300\times900$, $r-r_s\ge2\times10^{-5}$, Newton & main family: 1--3 pairs; from $\kappa\approx7.5$ second class\\
front, $\rho\in[0.10,0.26]$, $z\in[-1.50,-0.95]$ & $7.00$--$10.00$ (step $0.05$) & as above & rectangular, step $4\times10^{-4}$, $r-r_s\ge10^{-4}$ & quartets and bulk pairs, up to 2 quartets\\
front, $\rho\in[0.02,0.26]$, $z\in[-1.50,-0.95]$ & $8.20$--$8.45$ (step $0.01$) & as above & rectangular, step $4\times10^{-4}$, $r-r_s\ge10^{-4}$ & one second-class pair: birth, quartet, end in the bulk\\
control & $6$, $8$, $10$ & $4\kappa+30$ & Newton from scan & $|\Delta|=0$ in floating point\\
rear, $z\in[0,50]$, $\rho\le10$ & $3$; $2.5$ & radial functions to $x=52$; $200/185$ ($\kappa=3$), $170/155$ ($\kappa=2.5$) & $426\times1667$, Newton & none (both $l_\mathrm{max}$); $\min|S_\mathrm{pol}|=9.5\times10^{-3}$\\
front far, $z\in[-20,-5]$, $\rho\le10$ & $3$ & as above, $200/185$ & $426\times501$ & none; $\min|S_\mathrm{pol}|=0.47$\\
\end{tabular}
\end{ruledtabular}
\end{table}

\textit{Far region.} The rear and far-front grids were built with radial functions scaled for large $l$ (the first
attempt overflowed for $l\gtrsim150$). This model agrees with the near-field model to $10^{-8}$ in the radial functions,
$5\times10^{-9}$ in $\avS$, and reproduces O and X to $3\times10^{-10}$. Physical checks: far in front
$|\avS|=0.498$ ($\frac12$ expected); on the rear axis $|\avS|=9.50$--$9.98$ at $z=10$--$50$ for $\kappa=3$, against the
weak-field wave-lensing value $\frac12\pi w/(1-e^{-\pi w})=9.43$, $w=2\kappa$. Grids of spacing $0.008\times0.03$ do not
exclude pairs smaller than $\sim0.01$, and the rear was not scanned at other $\kappa$.

\textit{Main family, $6\le\kappa\le10$.} O is an L circle, X is not. Births (zeros of $a(z)$ on the axis):
$\kappa^*=5.6246$, $6.1232$, $6.5703$, $6.9960$, $7.4137$, $7.8290$, $8.2455$, $8.6722$, $9.1922$, $9.6346$, at
$r-r_s\approx0.40$--$0.57$. Deaths: $6.2653$, $6.7877$, $7.3183$, $7.8498$, $8.3716$, $8.9056$, $9.4322$, $9.9566$
(three estimates, $r_O-r_s\to0$ quadratically and linearly and $|\mathrm O-\mathrm X|\to0$, agree to $2\times10^{-4}$).
Lifetimes grow from $0.64$ to $1.28$; one pair coexists to $\kappa\approx6.1$, then one or two, and from
$\kappa\approx7.4$ up to three. Further zeros of $a$ at $r\approx1.003$--$1.006$ ($\kappa=7.009$, $9.776$, $9.909$,
$9.952$) lie at the lower edge of the grid, have no counterpart in the scan, and are left undecided.

\textit{Second class ($\kappa\gtrsim7.5$).} At $\rho\approx0.14$--$0.22$, farther from the axis than the main family, an
O--X pair whose members are both non-L circles is born in the bulk at $r-r_s\approx0.18$--$0.37$ (saddle-center
bifurcation; seen, e.g., at $\kappa=8.25$, $9.10$, $9.30$, $9.50$, $9.90$) and turns into a quartet of two O circles and
two L saddles. The two O circles rotate in opposite senses (vorticity $\partial_\rho S_z-\partial_zS_\rho$ of opposite sign),
so no saddle is needed between them; the two L saddles lie on opposite sides of the segment O--O. (The earlier notation
X$_\mathrm L$--O--O--X$_\mathrm L$ listed the circles by distance from the horizon, not along a line.) The quartet
($\Delta r\approx10^{-3}$--$10^{-2}$, $\Delta\rho\approx0.02$--$0.04$) drifts toward the horizon and ends either there
($\kappa\approx7.58$, $8.65$, $9.16$, $9.70$; these exits are zeros of $\gamma_H$, Sec.~\ref{sec:phase}) or in the bulk.
Up to two quartets coexist ($\kappa=9.05$, $9.45$, $9.85$), and with the main family six to seven O--X pairs are present
at $\kappa\approx9.4$--$9.9$. Quartets can also be born close to the horizon: on $[7.95,8.15]$ (step $0.01$, and $0.002$
on $[8.100,8.124]$, grids down to $r-r_s=10^{-5}$) the quartet seen at $7.95$ shrinks to one O--X pair at $8.00$ and
ends in the bulk in $(8.00,8.01]$ at $r-r_s\approx0.05$, and a second quartet lives only from $(8.108,8.110]$ to
$8.118$, at $r-r_s<3\times10^{-3}$, $\theta'\approx0.18$--$0.19$, and leaves through the horizon at the zeros of
$\gamma_H$ (Sec.~\ref{sec:phase}). The scan of step $0.05$ missed it entirely.

\textit{One window at step $0.01$} ($\kappa\in[8.20,8.45]$, rectangular grid as in Table~\ref{SM-tab:census}): the pair
is born in the bulk in $(8.23,8.24]$ at $r-r_s=0.181$, $\rho=0.203$; in $(8.25,8.26]$ a second O (vorticity $+1.4$,
against $-9.3$ for the first) and two L saddles appear, and the quartet persists to $8.36$, moving from $r-r_s=0.17$ to
$0.12$ with the first O at vorticity $-6$ to $-11$; at $8.36$ the outer O is an L circle next to a non-L saddle, and in
$(8.36,8.37]$ all four circles disappear in the bulk at $r-r_s\approx0.12$. Both transitions are faster than $0.01$.
Inside a quartet each O is the center of nested tori, but its core is not an L circle, so $q$ tends to a finite value
$q_0=\omega\rho_O/|S_\varphi^+(\mathrm O)|$ at the core, and the saddles, being L circles, have $S^+_\varphi(\mathrm X)=0$, so
$q$ need not tend to $0$ at the separatrix; the statement that every torus knot occurs therefore does not apply to these
cells (margins and $q$ in Table~\ref{tab:margins}).
\emph{Caveat:} outside this window the step $0.05$ does not resolve the events of the second class, so their $\kappa$ are
given only as intervals. The dense grid and the polar grid agree on all quartets; the dense grid added one or two
circles at eight values of $\kappa$ (pairs just after birth) and missed one dying circle below its cutoff
$r-r_s\ge10^{-4}$.

\section{Proper lengths}
\label{sec:proper}
With the spatial metric $dl^2=dr^2/f+r^2d\Omega^2$ the proper radial distance from the horizon is
$s(r)=\int_1^r dr/\sqrt f=\sqrt{r(r-1)}+\ln(\sqrt r+\sqrt{r-1})$ (quadrature agrees to $10^{-10}$), with
$s\approx2\sqrt{r-1}$ near the horizon. Radial lengths are stretched by $1/\sqrt f$; tangential lengths $r\,d\theta$ and
the circumference $2\pi R$ of a circle about the axis are proper. Table~\ref{tab:proper} lists the cells of the dense
scan ($\kappa\le6$). At the death of a cell its tangential size decreases as $r-r_s$ and its proper distance from the
horizon as $2\sqrt{r-r_s}$; the self-similarity $d\propto r-r_s$ is a statement about coordinates.

\begin{table}[ht]
\caption{Coordinate and proper sizes of the cells. $\Delta r$, $\Delta s$: coordinate and proper radial extent of the
homoclinic loop; $r\Delta\theta$: tangential extent (proper); last two columns: loop perimeter and cross-section area,
coordinate / proper. Young second pairs at $\kappa=5.25$, $5.75$. The column $\Delta s/(2\sqrt{r_O-1})$ equals
$\Delta r/[2(r_O-1)]$ to leading order; it is constant along the death of one cell (self-similarity) but differs
between rows, which are cells at different stages of their lives.}
\label{tab:proper}
\begin{ruledtabular}
\begin{tabular}{ccccccccc}
$\kappa$ & $r_O-1$ & $s(O)$ & $x_*(O)$ & $2\pi R_O$ & $\Delta r$ / $\Delta s$ & $\frac{\Delta s}{2\sqrt{r_O-1}}$ & $r\Delta\theta$ & perimeter; area\\
\hline
2.00 & 0.02421 & 0.3124 & $-2.697$ & 1.171 & $2.5\times10^{-4}$ / $1.6\times10^{-3}$ & 0.0052 & $5.9\times10^{-3}$ & 0.0118/0.0126; $1.0\times10^{-6}$/$6.6\times10^{-6}$\\
2.50 & 0.06557 & 0.5177 & $-1.659$ & 1.102 & $1.1\times10^{-3}$ / $4.6\times10^{-3}$ & 0.0089 & $1.5\times10^{-2}$ & 0.0300/0.0324; $1.2\times10^{-5}$/$4.7\times10^{-5}$\\
3.00 & 0.09026 & 0.6098 & $-1.315$ & 0.896 & $1.3\times10^{-3}$ / $4.7\times10^{-3}$ & 0.0078 & $1.6\times10^{-2}$ & 0.0331/0.0353; $1.5\times10^{-5}$/$5.3\times10^{-5}$\\
3.50 & 0.11616 & 0.6946 & $-1.037$ & 0.807 & $1.8\times10^{-3}$ / $5.4\times10^{-3}$ & 0.0080 & $1.9\times10^{-2}$ & 0.0383/0.0408; $2.3\times10^{-5}$/$7.1\times10^{-5}$\\
4.00 & 0.13046 & 0.7378 & $-0.906$ & 0.845 & $2.7\times10^{-3}$ / $7.9\times10^{-3}$ & 0.0109 & $2.3\times10^{-2}$ & 0.0459/0.0498; $4.2\times10^{-5}$/$1.2\times10^{-4}$\\
4.50 & 0.13561 & 0.7528 & $-0.862$ & 0.774 & $2.7\times10^{-3}$ / $7.7\times10^{-3}$ & 0.0105 & $2.2\times10^{-2}$ & 0.0438/0.0477; $4.0\times10^{-5}$/$1.2\times10^{-4}$\\
5.00 & 0.14598 & 0.7824 & $-0.778$ & 0.687 & $2.5\times10^{-3}$ / $7.0\times10^{-3}$ & 0.0092 & $2.0\times10^{-2}$ & 0.0415/0.0450; $3.6\times10^{-5}$/$9.9\times10^{-5}$\\
5.25 & 0.33687 & 1.2230 & $+0.249$ & 0.468 & $2.4\times10^{-3}$ / $4.9\times10^{-3}$ & 0.0042 & $3.0\times10^{-2}$ & 0.0597/0.0609; $5.0\times10^{-5}$/$1.0\times10^{-4}$\\
5.50 & 0.15506 & 0.8075 & $-0.709$ & 0.696 & $3.2\times10^{-3}$ / $8.7\times10^{-3}$ & 0.0110 & $2.2\times10^{-2}$ & 0.0454/0.0499; $4.9\times10^{-5}$/$1.3\times10^{-4}$\\
5.75 & 0.32740 & 1.2041 & $+0.211$ & 0.413 & $2.0\times10^{-3}$ / $4.1\times10^{-3}$ & 0.0036 & $2.5\times10^{-2}$ & 0.0512/0.0523; $3.6\times10^{-5}$/$7.3\times10^{-5}$\\
6.00 & 0.15654 & 0.8115 & $-0.698$ & 0.678 & $3.4\times10^{-3}$ / $9.3\times10^{-3}$ & 0.0118 & $2.2\times10^{-2}$ & 0.0453/0.0503; $5.3\times10^{-5}$/$1.4\times10^{-4}$\\
\end{tabular}
\end{ruledtabular}
\end{table}

\section{\texorpdfstring{Rotation number, the layer $q<1$, and nondegeneracy margins}{Rotation number, the q<1 layer, margins}}
\label{sec:q}
On a torus $\psi=c$ inside a cell: poloidal period $T(\psi)=\oint dl/|\mathbf S_\mathrm{pol}|$, toroidal advance per
poloidal turn $\Delta\varphi(\psi)=\oint S_\varphi^+/(\rho|\mathbf S_\mathrm{pol}|)\,dl$ ($P=1$), and $q=2\pi/\Delta\varphi$,
the number of poloidal turns per toroidal turn (the inverse of the usual winding number and of the plasma safety
factor); $q(\psi;P)=q(\psi;1)/P$. Coarea form, with $u=\pm(\psi-\psi_O)>0$:
\begin{equation}
T(u)=\frac{d}{du}\iint_{u'<u}\rho\,d\rho\,dz,\qquad \Delta\varphi(u)=\frac{d}{du}\iint_{u'<u}S_\varphi^+\,d\rho\,dz .
\end{equation}
Asymptotics: (i) non-L core: $q\to q_0=\omega\rho_O/S_\varphi^+(\mathrm O)$; (ii) L core ($S_\varphi^+(\mathrm O)=0$):
$\Delta\varphi=Cu+O(u^2)$, $q\to\infty$; (iii) homoclinic separatrix of a saddle with eigenvalue $\lambda_X$:
$T\simeq\lambda_X^{-1}\ln(1/\delta)$, $\Delta\varphi\simeq S_\varphi^+(\mathrm X)/(\rho_X\lambda_X)\ln(1/\delta)$, so $q\to0$ iff
$S_\varphi^+(\mathrm X)\ne0$. In the cells at the horizon the core asymptotics hold to $4.9\times10^{-4}$ (relative
deviation of $\Delta\varphi/u$ from $C$ at $u=10^{-3}u_X$), and the separatrix slopes $dT/d\ln\delta^{-1}=1/\lambda_X$,
$d\Delta\varphi/d\ln\delta^{-1}=S^+_\varphi(\mathrm X)/(\rho_X\lambda_X)$ to $3\times10^{-6}$--$4\times10^{-5}$.
Two numerical methods (one poloidal period with DOP853 on the model field; QUADPACK on a quintic spline of $\psi$)
agree to $|\Delta q/q|\le1.3\times10^{-9}$ at $\kappa=3$. In all $95$ cells examined ($88$ Buchdahl, $7$ horizon)
$\Delta\varphi$ has a constant sign, although $S^+_\varphi$ changes sign on every torus of the horizon cells.

\textit{Why $\Delta\varphi=O(u)$ at an L core.} Let $\xi=(\rho-\rho_O,z-z_O)$, $u=\frac12\xi^\mathsf{T}M\xi+\frac16T_{ijk}\xi_i\xi_j\xi_k+\dots$
($M>0$) and $S_\varphi^+=\mathbf g\cdot\xi+\frac12\xi^\mathsf{T}H\xi+\dots$ (no constant term at an L core). In the coarea form,
$\Delta\varphi=\frac{d}{du}\iint_{u'<u}S_\varphi^+\,d\rho\,dz$. To leading order the region $u'<u$ is an ellipse centered
at O, over which the linear term $\mathbf g\cdot\xi$ integrates to zero; it contributes only through the cubic
distortion of the level set, at the same order as the Hessian term. Using the second and fourth moments of the ellipse,
\begin{equation}
\Delta\varphi=Cu+O(u^2),\qquad C=\frac{\pi}{\sqrt{\det M}}\Bigl[\mathrm{tr}(HM^{-1})-\mathbf g^\mathsf{T}M^{-1}\mathbf v\Bigr],\quad
v_i=T_{ijk}(M^{-1})_{jk}.
\end{equation}
At $\kappa=3$, with derivatives by finite differences, $C=1.2269\times10^6$ (Hessian part $-0.09\times10^6$, gradient part
$+1.32\times10^6$); the orbits give $\Delta\varphi/u=C(1+\epsilon)$ with $\epsilon=4.9\times10^{-3}$, $1.5\times10^{-3}$,
$4.9\times10^{-4}$, $1.4\times10^{-4}$ at $u/u_X=10^{-2}$, $3\times10^{-3}$, $10^{-3}$, $3\times10^{-4}$, i.e., $\epsilon\propto u$.

\textit{The layer $q<1$.} We use one definition, $h=|\psi-\psi_O|$, $h_X=|\psi_X-\psi_O|$,
\begin{equation}
\delta=1-h/h_X=|\psi-\psi_X|/|\psi_O-\psi_X| .
\end{equation}
Table~\ref{tab:qlayer} gives $q(\delta)$ at $\kappa=3$. The separatrix law $\Delta\varphi=a+b\ln(1/\delta)$ fitted on
$10^{-10}\le\delta\le10^{-6}$ gives $b=0.294096$ against $S_\varphi^+(\mathrm X)/(\rho_X\lambda_X)=0.294112$. The increment
of $\Delta\varphi$ per decade is $0.6772$ from $\delta=10^{-4}$ to $10^{-9}$ and degrades below $10^{-10}$, where the
determination of the level of $\psi$ hits its numerical limit. The law crosses $q=1$ at $\delta=5.6\times10^{-10}$ and
$q=2/3$ at $1.3\times10^{-14}$.

\begin{table}[ht]
\caption{Rotation number $q$ ($P=1$) near the separatrix at $\kappa=3$. Values below the field-error floor
$\delta\approx2.8\times10^{-9}$ (Table~\ref{tab:margins}) are properties of the model field only.}
\label{tab:qlayer}
\begin{ruledtabular}
\begin{tabular}{ccccccccc}
$\delta$ & $10^{-4}$ & $10^{-6}$ & $10^{-7}$ & $10^{-8}$ & $10^{-9}$ & $3\times10^{-10}$ & $10^{-10}$ & $10^{-11}$\\
$q$ & 2.305 & 1.540 & 1.321 & 1.156 & 1.028 & 0.972 & 0.925 & 0.856\\
\end{tabular}
\end{ruledtabular}
\end{table}

\textit{Status of the knot statement.} The hypotheses are: a single O--X pair in the cell, O nondegenerate
($\det J>0$), X hyperbolic ($\lambda_X>0$) with a homoclinic loop, $S_\varphi^+(\mathrm X)\ne0$, and an L core with
$|\deg\mathbf b|=1$. Here $\mathbf b$ is the minor semi-axis $\mathbf B$ of the polarization ellipse
$\Ep=e^{i\alpha}(\mathbf A+i\mathbf B)$ projected onto the plane perpendicular to $\mathbf A$; its zeros are exactly
the L points. For all $11$ cells of the scan ($\kappa=2$--$6$) the degree on circles of radius $0.02$, $0.1$ and
$0.3\,|\mathrm O-\mathrm X|$ around O is $\pm1$ (integer to $10^{-6}$), the zero of $\mathbf b$ coincides with O to
$\le2.9\times10^{-13}|\mathrm O-\mathrm X|$, and the degree around X is $0$. A nonzero degree makes the L point persist
under perturbations that preserve duality and axial symmetry; under duality breaking the core would become a generic
O circle with finite $q_0$. Given the hypotheses, $|\Delta\varphi(\psi)|$ is continuous on $(\psi_O,\psi_X)$ and tends
to $0$ at the core and to $\infty$ at the separatrix, so $q$ takes every value in $(0,\infty)$ (Darboux).
Table~\ref{tab:margins} compares the margins with the field error, estimated as the maximum over O, X and eight points
of the cell of three differences: $l_\mathrm{max}$ increased by $2\kappa+16$, spline and ODE versus direct evaluation,
and ODE tolerance $10^{-12}$. The partial-wave tail is negligible ($\le10^{-14}$); the error is dominated by the splines
and the ODE tolerance. Every tabulated margin exceeds the error at least $2.5\times10^5$ times (fourteen main-family
cells, $\kappa=2$--$9$). This is strong numerical support, not a certified transfer to the exact field: the margins are
floating-point estimates at O, X and eight points, not a uniform $C^1$ bound, and they do not cover the global
structure of the separatrix (a single O--X pair and no other zeros in the cell), which rests on the census grids.
Two further checks at $\kappa=3$: on the circles $0.02$, $0.1$, $0.3\,|\mathrm O-\mathrm X|$ used for the degree, the
model field and the partial-wave sum without splines give $\max|\delta\mathbf b|=3.7\times10^{-10}$ against
$\min|\mathbf b|=6.0\times10^{-4}$, $3.0\times10^{-3}$, $8.9\times10^{-3}$ (ratios $\ge1.6\times10^6$); on a uniform
$21\times21$ grid over the cell $|\delta\avS|\le4.4\times10^{-10}$, while $|S_\mathrm{pol}|\ge1.9\times10^{-4}$ outside
$0.1|\mathrm O-\mathrm X|$ of O and X. Specific $q<1$ values lie below the floor $\delta=|\delta\Delta\psi|/|\Delta\psi|$ and
are properties of the model field. The margins were computed with floating-point estimates, not interval arithmetic.

\begin{table}[ht]
\caption{Nondegeneracy margins of the cells versus the field error ($P=1$, $l_\mathrm{max}=7\kappa+22$). Min ratio: the
smallest of $|S_\varphi^+(\mathrm X)|/|\delta S|$, $\lambda_X/|\delta\nabla S|$, $\omega^2/(\|J\||\delta\nabla S|)$,
$|\mathrm O-\mathrm X|/\delta x$ ($\delta x=\|J^{-1}\||\delta S|$) and $|\Delta\psi|/|\delta\Delta\psi|$; in every case it is
$|\mathrm O-\mathrm X|/\delta x$. Floor: $|\delta\Delta\psi|/|\Delta\psi|$. In every case the nondegeneracy margin exceeds the field error
by at least $2.5\times10^5$. Rows $\kappa=7$--$9$ (main family, cell in mid-life): base model
$l_\mathrm{max}=\lfloor2.5\kappa\rfloor+20$, $r_\mathrm{max}=3$, compared with $4\kappa+30$, the sum without splines, and
rtol $10^{-12}$; $\min|\mathbf b|$ on the circle $0.1|\mathrm O-\mathrm X|$; the degree of $\mathbf b$ around O is $-1$ on
circles $0.02$, $0.1$, $0.3\,|\mathrm O-\mathrm X|$ and $0$ around X. Last two rows: the two cells of a second-class
quartet, each O with the L saddle nearest in $\psi$; $^{b}$minimum over the other margins, since $S^+_\varphi(\mathrm X)=0$
exactly; $q\in[3.05,3.16]$ and $[7.60,8.07]$ from core to $\delta=0.01$.}
\label{tab:margins}
\begin{ruledtabular}
\begin{tabular}{cccccccccc}
$\kappa$ & $|\mathrm O-\mathrm X|$ & $\det J(\mathrm O)$ & $\lambda_X$ & $S_\varphi^+(\mathrm X)$ & $\min|\mathbf b|$ & $|\delta S|$ & $|\delta\nabla S|$ & floor & min ratio\\
\hline
2.00 & $3.9\times10^{-3}$ & 0.071 & 0.262 & $7.3\times10^{-3}$ & $7.5\times10^{-4}$ & $6.6\times10^{-11}$ & $3.7\times10^{-9}$ & $2.2\times10^{-9}$ & $8.6\times10^5$\\
2.50 & $1.0\times10^{-2}$ & 0.490 & 0.671 & $2.8\times10^{-2}$ & $2.5\times10^{-3}$ & $2.2\times10^{-10}$ & $6.9\times10^{-9}$ & $2.2\times10^{-9}$ & $3.1\times10^6$\\
3.00 & $1.1\times10^{-2}$ & 0.709 & 0.792 & $3.1\times10^{-2}$ & $3.0\times10^{-3}$ & $3.9\times10^{-10}$ & $1.2\times10^{-8}$ & $2.8\times10^{-9}$ & $2.4\times10^6$\\
3.50 & $1.3\times10^{-2}$ & 1.33 & 1.07 & $4.1\times10^{-2}$ & $3.9\times10^{-3}$ & $8.2\times10^{-10}$ & $2.7\times10^{-8}$ & $1.4\times10^{-8}$ & $2.1\times10^6$\\
4.00 & $1.5\times10^{-2}$ & 4.41 & 1.94 & $8.5\times10^{-2}$ & $6.5\times10^{-3}$ & $2.8\times10^{-9}$ & $5.1\times10^{-8}$ & $1.3\times10^{-8}$ & $1.7\times10^6$\\
4.50 & $1.5\times10^{-2}$ & 6.52 & 2.36 & $9.6\times10^{-2}$ & $7.1\times10^{-3}$ & $4.3\times10^{-9}$ & $7.1\times10^{-8}$ & $1.1\times10^{-8}$ & $1.3\times10^6$\\
5.00 & $1.4\times10^{-2}$ & 7.40 & 2.49 & $9.2\times10^{-2}$ & $7.1\times10^{-3}$ & $6.8\times10^{-9}$ & $1.0\times10^{-7}$ & $2.1\times10^{-8}$ & $8.4\times10^5$\\
5.25 & $2.1\times10^{-2}$ & 1.32 & 0.854 & $2.3\times10^{-2}$ & $4.0\times10^{-3}$ & $5.4\times10^{-9}$ & $1.4\times10^{-7}$ & $3.2\times10^{-8}$ & $3.8\times10^5$\\
5.50 & $1.5\times10^{-2}$ & 16.1 & 3.69 & 0.145 & $9.6\times10^{-3}$ & $1.2\times10^{-8}$ & $1.8\times10^{-7}$ & $2.1\times10^{-8}$ & $9.0\times10^5$\\
5.75 & $1.8\times10^{-2}$ & 1.39 & 0.886 & $2.1\times10^{-2}$ & $3.8\times10^{-3}$ & $7.1\times10^{-9}$ & $2.1\times10^{-7}$ & $6.6\times10^{-8}$ & $2.5\times10^5$\\
6.00 & $1.5\times10^{-2}$ & 27.3 & 4.84 & 0.190 & $1.1\times10^{-2}$ & $1.6\times10^{-8}$ & $2.5\times10^{-7}$ & $2.3\times10^{-8}$ & $9.2\times10^5$\\
\hline
7.00 & $1.4\times10^{-2}$ & 39.9 & 5.82 & 0.203 & $1.2\times10^{-2}$ & $2.7\times10^{-8}$ & $4.9\times10^{-7}$ & --- & $6.5\times10^5$\\
8.00 & $1.3\times10^{-2}$ & 73.9 & 7.99 & 0.266 & $1.4\times10^{-2}$ & $4.3\times10^{-8}$ & $8.0\times10^{-7}$ & --- & $5.8\times10^5$\\
9.00 & $1.3\times10^{-2}$ & 144 & 11.3 & 0.373 & $1.8\times10^{-2}$ & $6.4\times10^{-8}$ & $1.3\times10^{-6}$ & --- & $6.1\times10^5$\\
\hline
8.30, O$^+$ & $1.6\times10^{-2}$ & 9.25 & 2.11 & $0$ (L) & n/a & $1.2\times10^{-8}$ & $7.1\times10^{-7}$ & --- & $1.2\times10^6$$^{b}$\\
8.30, O$^-$ & $3.8\times10^{-3}$ & 1.75 & 1.47 & $0$ (L) & n/a & $4.6\times10^{-9}$ & $6.9\times10^{-7}$ & --- & $5.2\times10^5$$^{b}$\\
\end{tabular}
\end{ruledtabular}
\end{table}
The smallest margins belong to the young second pairs at $\kappa=5.25$ and $5.75$ ($\det J\approx1.3$). At a birth or
death the margins vanish by definition. The sampled frequencies are not uniformly far from these events: the distance
from the tabulated cell's own birth or death ranges from $0.015$ ($\kappa=2$, death at $2.015$) and $0.046$
($\kappa=2.5$) to $0.26$, and the nearest event of any pair is $0.015$--$0.19$ away (e.g., $0.020$ at $\kappa=5.75$,
from the death at $5.730$). Even the sample $0.015$ from a death has a margin ratio $8.6\times10^5$.

\section{Construction of knots and invariants}
\label{sec:knots}
Because the flow is axially symmetric, the first-return map on a torus is a rigid rotation by $\Delta\varphi$. A closed
line with $q=m/n$ is obtained from one poloidal orbit $(\rho,z,\varphi)(\tau)$ rotated $m$ times; the target level is
found by root finding on $\ln\delta$. Independent test: full 3D integration (DOP853, rtol $10^{-12}$) over $m$
periods. Invariants: linking number with the core ($=m$) and with the axis ($=-n$; the axis is closed through
infinity). On a torus of revolution these two numbers already determine $(m,n)$, so the Alexander polynomial
$\Delta(t)=(t^{mn}-1)(t-1)/[(t^m-1)(t^n-1)]$ is a consistency check. It was computed after a similarity rescaling of the
curve to size $\sim100$, since the knot-invariant library loses crossings for curves of size $\sim0.1$. For the
figure, the meridional section is enlarged by $(\rho,z)\mapsto\mathrm O+a(\rho-\rho_O,z-z_O)$, an isotopy away from the
axis.

\begin{table}[ht]
\caption{Closed flow lines of $\avS$ ($P=1$ unless stated). $\delta=1-h/h_X$: level of the torus relative to the
separatrix; floor: whether $\delta$ lies above the field-error floor of Table~\ref{tab:margins} (computed for the
horizon cells only); closure: minimum distance from the starting point after $m$ poloidal periods in full 3D
integration, in units of the torus radius (PEC: in units of $r_s$). In every row the linking numbers are $m$ and $-n$
(integer to $\le10^{-6}$) and the Alexander polynomial equals that of $T(m,n)$.}
\label{tab:knots}
\begin{ruledtabular}
\begin{tabular}{llcccc}
cell & $q=m/n$ & knot & $\delta$ & floor & closure\\
\hline
horizon, $\kappa=3$ & $3/2$ & $T(2,3)$ & $6.9\times10^{-7}$ & above & $3.8\times10^{-5}$\\
 & $5/2$ & $T(2,5)$ & $2.1\times10^{-4}$ & above & $1.7\times10^{-6}$\\
 & $5/3$ & $T(3,5)$ & $2.9\times10^{-6}$ & above & $2.5\times10^{-5}$\\
 & $8/5$ & $T(5,8)$ & $1.7\times10^{-6}$ & above & $9.6\times10^{-5}$\\
 & $7/2$ & $T(2,7)$ & $2.4\times10^{-3}$ & above & $2.4\times10^{-6}$\\
 & $4/3$ & $T(3,4)$ & $1.2\times10^{-7}$ & above & $2.4\times10^{-4}$\\
 & $2/3$, $3/5$, $5/8$, $2/5$ & as $3/2$, $5/3$, $8/5$, $5/2$ & $\le1.3\times10^{-14}$ & below & not constructed\\
Buchdahl, $\kappa=1$ & $2/3$ & $T(2,3)$ & $9.3\times10^{-7}$ & n/a & $3.3\times10^{-4}$\\
 & $3/2$ & $T(2,3)$ & $6.2\times10^{-3}$ & n/a & $2.4\times10^{-7}$\\
 & $3/5$ & $T(3,5)$ & $1.6\times10^{-7}$ & n/a & $7.1\times10^{-4}$\\
 & $5/8$ & $T(5,8)$ & $3.2\times10^{-7}$ & n/a & $4.1\times10^{-3}$\\
 & $5/2$ & $T(2,5)$ & $8.0\times10^{-2}$ & n/a & $8\times10^{-7}$\\
 & $5/3$ & $T(3,5)$ & $1.2\times10^{-2}$ & n/a & $1.5\times10^{-7}$\\
 & $7/2$ & $T(2,7)$ & $0.19$ & n/a & $4.5\times10^{-7}$\\
 & $3/2$, $P=0.5$ & $T(2,3)$ & $5.4\times10^{-6}$ & n/a & $7.4\times10^{-6}$\\
PEC, $\kappa=3$, cell \#3 & $3/2$ & $T(2,3)$ & $0.456$ & n/a & $2.4\times10^{-10}r_s$\\
 & $2/3$ & $T(2,3)$ & $0.065$ & n/a & $3.0\times10^{-9}r_s$\\
\end{tabular}
\end{ruledtabular}
\end{table}
For $q<1$ the knot types repeat those of $q>1$ ($T(m,n)\cong T(n,m)$), with the same chirality. These knots are
not the knotted field lines (hopfions) of Refs.~\cite{Ranada1989,IrvineBouwmeester2008,Kedia2013}, nor knotted optical
vortices~\cite{Dennis2010}: those are lines of the instantaneous fields (or phase singularities) and evolve in time,
while ours are stationary streamlines of $\avS$. For a pure helicity the instantaneous field here simply rotates
rigidly about the axis.
The Buchdahl cell of Table~\ref{tab:knots} lies inside the star, with its core at $(\rho,z)=(0.568,-0.007)$; the PEC
cell has $q\in[0.093,2.29]$ at $P=1$. Convention: $q=m/n$ with $\gcd(m,n)=1$; the closed line makes $m$ poloidal
turns (linking number $m$ with the core) and $n$ toroidal turns (linking number $-n$ with the axis) and is the torus
knot $T(m,n)\cong T(n,m)$. The table lists each knot as $T(\min,\max)$. Levels $q>1$ lie toward the core and give
every type; levels $q<1$ lie in the layer at the separatrix, and at $\kappa=3$ they are below the field-error floor.
Reversing the helicity ($P\to-P$) reverses $\Delta\varphi$ and hence the chirality of every knot.

\section{Watson--Sommerfeld transformation for the non-alternating near-field sum}
\label{sec:ws}
On the front axis ($\theta=\pi$), with $P_l(-1)=(-1)^l$, TM coefficient $a_l=-i^l(2l+1)/(\kappa L)$ and radial
normalization $R_l=-(1/2i)\,i^lu_\mathrm{hor}/A_\mathrm{in}$, the phase factors combine to $(-1)^l i^l i^l=1$:
\begin{equation}
a(r)=\sum_{l\ge1}h(l+\tfrac12),\qquad h(\lambda)=\frac{2\lambda\,w(r;\lambda)}{A_\mathrm{in}(\lambda)},\qquad
w=\frac{u_\mathrm{hor}/r-fu'_\mathrm{hor}/(i\kappa r)}{4i\kappa}.
\end{equation}
$h$ is odd in $\lambda$ and analytic except at $\pm\lambda_n$. With $-\pi\tan\pi\lambda$ (residues 1 at $\lambda=l+\frac12$)
and the splitting $\tan\pi\lambda=\pm i\mp ie^{\pm i\pi\lambda}/\cos\pi\lambda$ on the upper/lower branches,
\begin{align}
a(r)&=B(r)+\sum_nP_n(r),\\
B&=\int_0^\infty h\,d\lambda+\int_0^\infty2y\,g(-y^2-\tfrac14)\frac{e^{-\pi y}}{\cosh\pi y}\,dy-h(\tfrac12),\quad g=w/A_\mathrm{in},\\
P_n&=-\pi i\,\mathrm{Res}_{\lambda_n}h\,K(\lambda_n),\qquad K(\lambda)=\frac{e^{i\pi\lambda}}{\cos\pi\lambda}=\frac{2e^{2\pi i\lambda}}{1+e^{2\pi i\lambda}} .
\end{align}
In the second integral $g$ is evaluated on the imaginary $\lambda$ axis, $\lambda=iy$, as a function of
$L=\lambda^2-\frac14=-y^2-\frac14$. The term $-h(\frac12)$ removes the nonexistent $l=0$ photon mode.

\textit{Assumptions, checked numerically (not proven).} Only the remainders carrying $K$ are deformed: $hK$ into the
first quadrant (it collects the poles) and $h\bar K$, $\bar K=e^{-2\pi i\lambda}/(1+e^{-2\pi i\lambda})$, into the
fourth (it gives the imaginary-axis integral); the integral of $h$ itself stays on the real axis. (i) Arcs: on
$|\lambda|=R$ in the first quadrant, $\max|h_HK|=2.8\times10^{-3}$, $7.4\times10^{-11}$, $3.7\times10^{-19}$ for $R=10$,
$15$, $20$ at $\kappa=3$, and $4.0\times10^{-6}$, $2.1\times10^{-13}$, $5.9\times10^{-29}$ for $R=25$, $30$, $40$ at
$\kappa=8$; in the fourth quadrant $\max|h_H\bar K|$ falls as $e^{-4.1R}$ at $\kappa=3$ ($R\ge15$) and is $2\times10^{-6}$,
$1\times10^{-13}$ at $R=25$, $30$ for $\kappa=8$. For $R<b_c\kappa$ the arc still meets the real axis below the barrier
and the integrand is $O(R/\kappa)$; the decay starts beyond $b_c\kappa$, where the waves tunnel. The same holds for
$h(\lambda;r=1.1)$. On the imaginary axis $|h_H(iy)|e^{-\pi y}/\cosh\pi y\le5\times10^{-4}$. (ii) No poles in the
fourth quadrant: the winding number of $A_\mathrm{in}$ around the rectangle $\mathrm{Re}\lambda\in[0.05,30]$,
$\mathrm{Im}\lambda\in[-4,-0.02]$ is $0$ at $\kappa=3$ and $8$ (step $0.01$, largest phase step $0.06$~rad), and
$|A_\mathrm{in}|\ge1$ on a $0.1\times0.1$ grid inside. (iii) The pole series converges,
$|K(\lambda_n)|\approx2e^{-2\pi(n+1/2)}$.

\textit{Accuracy.} On the front axis, background plus five poles reproduces the partial-wave sum to $7.5\times10^{-5}$
(maximum, at $z=-5$, set by the truncation of the real-axis integral) and $2.6\times10^{-8}$ (median) of the incident
amplitude for $\kappa\in[1.5,5]$, $z\in[-5,-1.004]$; with one, two and three poles the maxima are $0.21$,
$1.1\times10^{-2}$ and $4.1\times10^{-4}$. At the horizon, $B_H+P_0+\dots+P_3$ reproduces $a_H$ to $1.8\times10^{-4}$ for
$\kappa\le9.55$. The membrane $\bar f$ is a control for the cell but not for the CAM decomposition: its $n=2$ pole does
not converge to the horizon pole in the range studied (Sec.~\ref{sec:controls}).

\textit{Off the axis.} With $\theta'=\pi-\theta$, $\mu'=\cos\theta'$, $\nu=\lambda-\frac12$,
$P_\nu(\mu')={}_2F_1(-\nu,\nu+1;1;\sin^2\frac{\theta'}{2})$ is entire in $\nu$ and
$P_\nu'(\mu')=\frac{L}{2}G_1$, $G_1={}_2F_1(1-\nu,\nu+2;2;\sin^2\frac{\theta'}{2})$. The factor $1/L$ of the partial-wave
coefficient cancels, and the field of helicity $+$ at $\varphi=0$ is built from five sums
\begin{equation}
\mathcal A_j=\sum_{l\ge1}\hat c_l\,F_j,\qquad \hat c_l=\frac{\lambda}{i\kappa A_\mathrm{in}},\qquad
(F_1,\dots,F_5)=\Bigl(u\sin\theta'\,\tfrac{L}{2}G_1,\ u\,\tfrac{G_1}{2},\ u\,\bigl(P_\nu-\mu'\tfrac{G_1}{2}\bigr),\
u'\,\tfrac{G_1}{2},\ u'\,\bigl(P_\nu-\mu'\tfrac{G_1}{2}\bigr)\Bigr),
\end{equation}
as $E_r=\mathcal A_1/(i\kappa r^2\sqrt2)$, $E_\theta=[\mathcal A_2/r-f\mathcal A_5/(i\kappa r)]/\sqrt2$,
$E_\varphi=-i[\mathcal A_3/r-f\mathcal A_4/(i\kappa r)]/\sqrt2$, and $\mathbf H=-i\mathbf E$. Every summand is analytic and
even in $\lambda$ times $\lambda$, and all five sums are non-alternating near the front pole, so each is decomposed as
on the axis, $\mathcal A_j=B_j+\sum_nP_{n,j}$, with the same kernel and the same poles; no travelling-wave decomposition
is needed near the front pole. The flow then follows from $\avS=\frac12\mathrm{Re}(\mathbf E\times\mathbf H^*)$.
Background alone and poles alone ($n\le4$) produce no zero of $\avS$ near the cell at $\kappa=3$; background plus
$P_0$ produces the pair shifted by $5.4\times10^{-3}r_s$ with $\psi_O-\psi_X$ off by $+35\%$; with $P_1$ the shift is
$3.2\times10^{-4}r_s$ ($+0.8\%$), with $P_0..P_2$ $8\times10^{-6}$, and with $P_0..P_4$ $1\times10^{-9}r_s$.
\textit{Local test at $\kappa=3$} (O, X and an $11\times11$ grid in the cell): the error of $(S_\rho,S_z)$ at O is
$4.2\times10^{-2}$, $6.2\times10^{-4}$, $6.4\times10^{-6}$, $3.8\times10^{-7}$, $8.3\times10^{-9}$ with $1$--$5$ poles, that of
its gradient $5.3\times10^{-1}$, $7.6\times10^{-3}$, $2.1\times10^{-4}$, $9.7\times10^{-6}$, $2.6\times10^{-7}$ (scale $7.7$),
with the same numbers at X; on the grid the error relative to $\max|\avS|$ falls from $2.5\times10^{-2}$ (two poles) to
$2.2\times10^{-7}$ (five). \textit{Higher frequencies:} on a grid $r-1\in[0.002,0.8]$, $\theta'\le0.4$, the five sums are
reproduced with five poles to $8.5\times10^{-7}$ ($\kappa=6.27$) and $2.1\times10^{-5}$ ($\kappa=8$; $2.3\times10^{-6}$ for
$r-1\le0.1$) of their maxima, with three poles to $6.6\times10^{-4}$.

\section{The death condition: near-horizon expansion, phase condition, and the constant $\pi/4$}
\label{sec:phase}
\textit{Near-horizon expansion and the exact death condition.} Near the horizon every partial wave has the same
leading behavior, $u_l=t^{-i\kappa}[1+O(t)]$ and $fu_l'=-i\kappa u_l[1+O(t)]$, $t=r-1$. In the sums of
Sec.~\ref{sec:ws} this gives $\mathcal A_4\to-i\kappa\mathcal A_2$, $\mathcal A_5\to-i\kappa\mathcal A_3$, hence, for
helicity $+$ at $\varphi=0$,
\begin{equation}
\mathbf E_t=\tfrac{1}{\sqrt2}\,t^{-i\kappa}\gamma_H(\theta')\,(\hat{\boldsymbol\theta}-i\hat{\boldsymbol\varphi})+O(t),\qquad
\gamma_H(\theta')=\sum_{l\ge1}\frac{\lambda}{i\kappa A_\mathrm{in}(\lambda)}\Bigl[P_l(\mu')+\frac{(1-\mu')P_l'(\mu')}{l(l+1)}\Bigr],
\label{eq:gammaH}
\end{equation}
$\mu'=\cos\theta'$: the tangential field at the horizon is circularly polarized, and $\gamma_H(0)=a_H$ below. With
$\mathbf H=-i\mathbf E$, $S_r=-\frac12|\gamma_H|^2+O(t)$ and $S_\theta=-\mathrm{Im}(E_\varphi E_r^*)$, which also vanishes
where $\gamma_H=0$. A zero of the meridional flow can therefore reach the horizon only at a zero of $\gamma_H$, so a cell
can die only at
\begin{equation}
\gamma_H(\theta'_d;\kappa_d)=0,
\label{eq:exactdeath}
\end{equation}
one complex equation for two real unknowns: zeros are isolated points of the $(\kappa,\theta')$ plane. Near such a
point $\gamma_H$ is linear in $\theta'-\theta'_d$ and $\kappa-\kappa_d$, and the $O(t)$ terms let a zero of the flow
exist at $t\propto|\kappa-\kappa_d|$ on one side, which is the observed linear approach $r_O-r_s\propto\kappa_d-\kappa$
with $|\mathrm O-\mathrm X|\propto t$ (this last step is a scaling argument, not a proof). Equation~\eqref{eq:exactdeath}
is thus a necessary condition, derived; that a dying pair exists at every zero (sufficiency) rests on this scaling
argument and is checked on the sixteen deaths and on the exits of quartets below. Checks: at $\kappa=3$ the
tangential field obeys Eq.~\eqref{eq:gammaH} with a relative error $7.3\,t$ for $t=10^{-3}$--$10^{-6}$, and
$|S_r+\frac12|\gamma_H|^2|/\max|\gamma_H|^2=7.7\,t$. We located all zeros of $\gamma_H$ for $\kappa\in[1.5,10.1]$,
$\theta'\in[0,1]$ (grid $0.02\times5\times10^{-4}$, $l_\mathrm{max}=\lfloor2.5\kappa\rfloor+20$, bicubic spline, Newton;
two zeros recomputed without the spline agree to $10^{-6}$). There are $26$ (Table~\ref{tab:gammaH}): sixteen are the
deaths of the main family, and ten form five close pairs of opposite orientation, consistent with the two
counter-rotating cells of a quartet (the relation between the orientation of a zero and the sense of rotation is not
derived). Four pairs ($\kappa\approx7.58$, $8.64$--$8.65$, $9.16$--$9.17$, $9.70$--$9.71$) match
the quartets that the scan of step $0.05$ sees ending in the horizon. The pair at $\kappa\approx8.116$--$8.118$ had no
counterpart in that scan; a scan of step $0.002$--$0.01$ down to $r-r_s=10^{-5}$ finds a short-lived quartet born
near the horizon in $(8.108,8.110]$ at $r-r_s\approx2.8\times10^{-3}$, whose two cells leave through the horizon at
$\kappa_d=8.11559$, $\theta'=0.180$ and $8.11809$, $0.188$ (linear extrapolation of $r_O-r_s$), against the zeros
$8.115634$, $0.1794$ and $8.118091$, $0.1878$ (Sec.~\ref{sec:census}). All five pairs are thus exits of second-class
cells.
On the axis $\gamma_H$ has no zero ($|\gamma_H(0)|\ge0.33$).

\begin{table}[ht]
\caption{Zeros of $\gamma_H(\theta';\kappa)$, Eq.~\eqref{eq:exactdeath}, for $\kappa\le10.1$. Main family: $m$ as in
Table~\ref{tab:deaths}; $\Delta$: zero minus measured death. Second class: pairs with opposite sign of
$\det\partial(\mathrm{Re}\gamma_H,\mathrm{Im}\gamma_H)/\partial(\kappa,\theta')$. $^{d}$Exits of a quartet born near the horizon in
$(8.108,8.110]$; measured $\kappa_d=8.11559$ and $8.11809$ (Sec.~\ref{sec:census}).}
\label{tab:gammaH}
\begin{ruledtabular}
\begin{tabular}{cccc|cccc}
$m$ & $\kappa_d$ & $\theta'_d$ & $\Delta$ & $m$ & $\kappa_d$ & $\theta'_d$ & $\Delta$\\
\hline
4 & 2.015238 & 0.2032 & $+2.0\times10^{-4}$ & 12 & 6.265242 & 0.1011 & $-5\times10^{-5}$\\
5 & 2.545714 & 0.1875 & $-1.9\times10^{-4}$ & 13 & 6.787676 & 0.0959 & $+3\times10^{-5}$\\
6 & 3.081842 & 0.1576 & $0$ & 14 & 7.318282 & 0.0876 & $0$\\
7 & 3.616355 & 0.1534 & $+1.5\times10^{-4}$ & 15 & 7.849787 & 0.0855 & $+1\times10^{-5}$\\
8 & 4.140555 & 0.1356 & $+3.3\times10^{-4}$ & 16 & 8.371632 & 0.0802 & $0$\\
9 & 4.676965 & 0.1233 & (used as $\kappa_d$) & 17 & 8.905635 & 0.0753 & $0$\\
10 & 5.203592 & 0.1187 & $0$ & 18 & 9.432220 & 0.0738 & $+7\times10^{-5}$\\
11 & 5.729985 & 0.1060 & $0$ & 19 & 9.956592 & 0.0689 & $0$\\
\hline
\multicolumn{8}{l}{second class: $(7.580733, 0.1845)$ and $(7.587015, 0.2071)$; $(8.115634, 0.1794)$ and $(8.118091, 0.1878)^{d}$;}\\
\multicolumn{8}{l}{$(8.638867, 0.1514)$ and $(8.650511, 0.1938)$; $(9.162234, 0.1448)$ and $(9.172567, 0.1797)$; $(9.696757, 0.1354)$ and $(9.707747, 0.1725)$}\\
\end{tabular}
\end{ruledtabular}
\end{table}

\textit{Horizon limit on the axis and the phase criteria.} On the axis ($\theta'=0$), $w(r;\lambda)\to t^{-i\kappa}/(2i\kappa)$
independently of $\lambda$ and $a(r)=t^{-i\kappa}a_H(\kappa)(1+O(t))$ with
\begin{gather}
a_H=\sum_{l\ge1}\frac{\lambda}{i\kappa A_\mathrm{in}(\lambda)}=B_H+\sum_nP_{n,H},\qquad
P_{n,H}=-\frac{\pi\lambda_nK(\lambda_n)}{\kappa A_\mathrm{in}'(\lambda_n)},\qquad B_H=\frac{I}{i\kappa},\\
I=\int_0^\infty\frac{\lambda\,d\lambda}{A_\mathrm{in}}+\int_0^\infty\frac{y\,e^{-\pi y}\,dy}{A_\mathrm{in}(iy)\cosh\pi y}-\frac{1}{2A_\mathrm{in}(\frac12)} .
\end{gather}
Off the axis the same transformation gives $\gamma_H(\theta')=B_H(\theta')+\sum_nP_{n,H}(\theta')$, with the factor
$P_\nu(\mu')+(1-\mu')G_1/2$ inserted in each integrand and residue. We define two families of criteria:
\begin{gather}
\Phi_H^{(N)}(\kappa)=\arg\sum_{n<N}P_{n,H}(0)-\arg[-B_H(0)]=0\bmod2\pi\quad\text{(on the axis)},\\
Z^{(N)}:\ B_H(\theta')+\sum_{n<N}P_{n,H}(\theta')=0\quad\text{(off the axis)}.
\end{gather}
$Z^{(N)}$ tends to Eq.~\eqref{eq:exactdeath} as $N\to\infty$; $\Phi_H^{(N)}$ uses only the phase and only the axis. The
single-pole phase decomposes exactly (to $6\times10^{-14}$~rad) as
\begin{equation}
\Phi_H^{(1)}=\frac{\pi}{2}+\arg\lambda_0+\arg K(\lambda_0)-\arg A_\mathrm{in}'(\lambda_0)-\arg I .
\end{equation}
$B_H$ plus four poles reproduces $a_H$ to $10^{-5}$ ($\kappa\le6$) and $1.8\times10^{-4}$ ($\kappa\le9.55$);
$A_\mathrm{in}'$ from Cauchy integrals on two radii agrees to $7\times10^{-6}$. Off the axis ($\theta'\le0.3$) $B_H+P_0+\dots
+P_4$ reproduces $\gamma_H$ to $8\times10^{-8}$--$3\times10^{-6}$ relative for $\kappa=2$--$9.4$.

\textit{Barrier top.} In the tortoise coordinate $u''+(\kappa^2-Lg)u=0$ with $g=f/x^2$, whose maximum is at $x=3/2$. Near it,
$\kappa^2-Lg\approx\kappa^2-L/b_c^2+k_2y^2$ with $y=x_*-x_*(3/2)$ and $k_2=16L/729$. In $s=cy$, $c=(4k_2)^{1/4}$, this is
Weber's equation $w''+(s^2/4-a)w=0$ with
\begin{equation}
a=\frac{L/b_c^2-\kappa^2}{2\sqrt{k_2}} .
\end{equation}
The solution ingoing at the horizon is $D_\nu(-e^{-i\pi/4}s)$, $\nu=-ia-\frac12$. The connection formula (Gradshteyn
9.248.1) gives its ingoing part at $s\to+\infty$. Matching to WKB solutions on both sides, with the phases
$\int p\,dx_*$ expanded as $p\approx\sqrt{Q_c}-\Delta L\,g/(2\sqrt{Q_c})$ (the logarithmic part gives $a\ln s$), yields
\begin{equation}
A_\mathrm{in}(\lambda)\simeq e^{i\kappa(1+D(r_s))}\,e^{ia(2\ln c+J)}\,e^{\pi a/2}\,\frac{\sqrt{2\pi}}{\Gamma(\frac12+ia)} .
\label{eq:ainSM}
\end{equation}
The factor $e^{i\kappa}$ in $e^{i\kappa(1+D)}$ comes from the normalization at the horizon,
$u_\mathrm{hor}=t^{-i\kappa}=e^{i\kappa}e^{-i\kappa x_*}$ as $x_*=r+\ln(r-1)\to1+\ln t$; $\kappa D(r_s)$ is the WKB phase deficit of the
critical wave. Checks: $|A_\mathrm{in}|^2=1+e^{2\pi a}$ (transmission through a parabolic barrier); the poles of $\Gamma$ at
$a=i(n+\frac12)$ give $\lambda_n\approx b_c\kappa+i(n+\frac12)$, with $da/d\lambda=27/(4b_c^2)=1$. The constants have
closed forms (substitution $u=\sqrt{1+3/x}$, using $1-b_c^2g=(2x-3)^2(x+3)/(4x^3)$):
\begin{align}
D(r_s)&=3\sqrt3-\tfrac92+6\ln\frac{2+\sqrt3}{2}=4.4390167209\quad(1.583214\ \text{inside},\ 2.855803\ \text{outside}\ x=3/2),\\
J&=6\ln3-3\ln(2+\sqrt3)=2.6408000\quad(J_L=3\ln3-2\ln(2+\sqrt3),\ J_R=3\ln3-\ln(2+\sqrt3)).
\end{align}
At $a=i/2$ the factor $e^{ia(2\ln c+J)}$ is real, $e^{\pi a/2}=e^{i\pi/4}$, and $d(1/\Gamma)/d\lambda=ia'$; hence
$\arg A_\mathrm{in}'(\lambda_0)=\kappa(1+D)+3\pi/4$. For the background, $1/A_\mathrm{in}\approx e^{-i\kappa-i\lambda^2/(2\kappa)}$
for $\lambda\ll\kappa$, so $I\approx-i\kappa e^{-i\kappa}$ and $-B_H\approx e^{-i\kappa}$. Together,
\begin{equation}
\Phi_H^{(1)}(\kappa)=2\pi\,\mathrm{Re}\,\lambda_0-\kappa D(r_s)+\frac\pi4-\arg\bigl(1+e^{2\pi i\lambda_0}\bigr)+\arg\lambda_0+O(1/\kappa).
\end{equation}
The constant $\pi/4$ is the parabolic-cylinder (Maslov-type) factor $e^{\pi a/2}=e^{i\pi/4}$, combined with $\pi/2$
from the derivative of $1/\Gamma$, $\pi/2$ from the factor $-\pi i$ and $-\pi/2$ from the background. The modulus also
has a closed form, $|A_\mathrm{in}'(\lambda_0)|\approx\sqrt{2\pi}e^{-J/2}/c$ (agreement $2.3\%$ at $\kappa=2$, $0.4\%$ at
$\kappa=5$), which gives
\begin{equation}
|P_{0,H}|\approx\frac{\pi|\lambda_0||K(\lambda_0)||c|e^{J/2}}{\kappa\sqrt{2\pi}}
\;\to\;\sqrt{2\pi}\,b_c\,e^{-\pi}\,\frac{27}{(2+\sqrt3)^{3/2}}\Bigl(\frac{4\kappa}{3\sqrt3}\Bigr)^{1/2}=0.9246917\sqrt\kappa .
\label{eq:P0H}
\end{equation}
The first form keeps the exact $\lambda_0$ and $|K(\lambda_0)|$, which oscillates by $\pm5\%$ about $2e^{-\pi}$; the
second replaces them by their leading values. At $\kappa=3$: exact $1.5949$, first form $1.6116$, second
$1.6016$; at $\kappa=8$: $2.5860$, $2.5898$, $2.6154$. $|B_H|=1.0009$--$1.0056$ for $\kappa=2$--$5$.

Table~\ref{tab:beta} compares the numerically exact components with their WKB limits; polynomial extrapolation in
$1/\kappa$ from $\kappa\ge6$ gives $2.356193$ ($3\pi/4=2.356194$), $-1.570796$ ($-\pi/2$), $\arg\lambda_0\to-10^{-6}$, and
$0.785399$ for the smooth part of $\beta$ ($\pi/4=0.785398$), with $1/\kappa$ coefficients $+0.147$, $-0.250$,
$+0.1925=1/(2b_c)$ and $+0.295$.

\begin{table}[ht]
\caption{Components of the horizon phase. $\beta=\Phi_H^{(1)}-[2\pi\mathrm{Re}\lambda_0-\kappa D(r_s)]$; WKB limits in brackets.}
\label{tab:beta}
\begin{ruledtabular}
\begin{tabular}{cccccc}
$\kappa$ & $\arg\lambda_0$ [0] & $-\arg(1+e^{2\pi i\lambda_0})$ & $\arg A_\mathrm{in}'-\kappa(1+D)$ [$3\pi/4$] & $\arg I+\kappa$ [$-\pi/2$] & $\beta$ [$\pi/4$ + osc.]\\
\hline
1.5 & 0.12234 & $+0.00790$ & 2.44615 & $-1.73722$ & 0.99210\\
2.0 & 0.09363 & $-0.04508$ & 2.42625 & $-1.69570$ & 0.88880\\
2.5 & 0.07563 & $+0.01179$ & 2.41325 & $-1.67075$ & 0.91572\\
3.0 & 0.06336 & $+0.03743$ & 2.40421 & $-1.65410$ & 0.92148\\
4.0 & 0.04778 & $-0.02161$ & 2.39257 & $-1.63328$ & 0.83768\\
5.0 & 0.03832 & $-0.00352$ & 2.38543 & $-1.62079$ & 0.84095\\
6.0 & 0.03198 & $+0.02814$ & 2.38062 & $-1.61246$ & 0.86275\\
8.0 & 0.02401 & $+0.04073$ & 2.37456 & $-1.60204$ & 0.86303\\
\end{tabular}
\end{ruledtabular}
\end{table}

\textit{Death condition: approximations.} Equation~(8) of the Letter,
$2\pi\mathrm{Re}\lambda_0-\arg(1+e^{2\pi i\lambda_0})+\arg\lambda_0-\kappa D(r_s)+\pi/4=2\pi m$, is the WKB form of
$\Phi_H^{(1)}=2\pi m$. Table~\ref{tab:deaths} compares it, the on-axis criteria $\Phi_H^{(1)}$, $\Phi_H^{(4)}$, a post hoc
variant (8$'$), and the eikonal formula of our earlier draft, $\kappa_d=(2\pi m-\beta_0)/(2\pi b_c-D(r_s))$ with a fitted
$\beta_0=-17.714$, with the sixteen measured deaths. The integer $m$ is the value of the left side of Eq.~(8) divided by
$2\pi$ at the measured death; that left side is continuous and increasing on $[1.5,10.2]$ without any unwrapping
($|e^{2\pi i\lambda_0}|<1$), and at the measured deaths it equals $2\pi(m+\delta)$ with $|\delta|\le0.022$. The earlier
version of this table numbered the predicted deaths $m=10$--$17$; the correct values are $m=12$--$19$. The deaths
$m=10$, $11$ belong to windows that overlap at $\kappa\approx5.2$ and $5.7$, were seen but not resolved in the scan of
$\kappa\le6$, and were measured in this revision, after the prediction test (pair tracked to $r_O-r_s=6.6\times10^{-5}$
and $2.4\times10^{-3}$; quadratic and linear extrapolation of $r_O-r_s$ and $|\mathrm O-\mathrm X|\to0$ agree to
$2\times10^{-5}$). For $m=9$ we use the zero of $\gamma_H$, $4.676965$, in place of the bisection value $4.675\pm0.025$.
The eight values in the column ``Eq.~(8)'' for $m=12$--$19$ are the numbers recorded before the
scan; relabeling does not change them.

\begin{table}[ht]
\caption{Measured deaths $\kappa_d$ of the main family and predicted minus measured values. Rows $m=4$--$9$: deaths
known before Eq.~(8) was written; rows $m=12$--$19$: deaths predicted before the scan of $[6,10]$ (column ``Eq.~(8)'':
values recorded before the scan). $^{a}$Measured in this revision, after the prediction test. $^{c}$Zero of $\gamma_H$ (Table~\ref{tab:gammaH}); bisection
in the scan of step $0.05$ gave $4.675\pm0.025$. $\Phi_H^{(N)}$ computed
to $\kappa=9.55$. (8$'$): Eq.~(8) plus $\arg(1+\rho_1)$, Eq.~\eqref{eq:rho1}, added post hoc. The zeros of $\gamma_H$
(Table~\ref{tab:gammaH}) agree with all measured values to $\le3.3\times10^{-4}$.}
\label{tab:deaths}
\begin{ruledtabular}
\begin{tabular}{cccccccc}
$m$ & $\kappa_d$ (measured) & Eq.~(8) & $\Delta$ Eq.~(8) & $\Delta\,\Phi_H^{(1)}$ & $\Delta\,\Phi_H^{(4)}$ & $\Delta$ (8$'$) & $\Delta$ eikonal (fitted)\\
\hline
4 & 2.01504 & 2.0147 & $-0.0004$ & $-0.0050$ & $-0.0083$ & $-0.0040$ & $+0.0040$\\
5 & 2.54590 & 2.5440 & $-0.0019$ & $-0.0053$ & $-0.0094$ & $-0.0063$ & $+0.0018$\\
6 & 3.08184 & 3.0820 & $+0.0001$ & $-0.0030$ & $-0.0087$ & $-0.0060$ & $-0.0054$\\
7 & 3.61621 & 3.6150 & $-0.0012$ & $-0.0035$ & $-0.0091$ & $-0.0070$ & $-0.0112$\\
8 & 4.14023 & 4.1410 & $+0.0008$ & $-0.0014$ & $-0.0085$ & $-0.0066$ & $-0.0065$\\
9$^{c}$ & 4.67697 & 4.6786 & $+0.0016$ & $-0.0003$ & $-0.0086$ & $-0.0070$ & $-0.0146$\\
\hline
10$^{a}$ & 5.20359 & 5.2046 & $+0.0010$ & $-0.0006$ & $-0.0084$ & $-0.0072$ & $-0.0126$\\
11$^{a}$ & 5.72998 & 5.7339 & $+0.0039$ & $+0.0023$ & $-0.0081$ & $-0.0069$ & $-0.0103$\\
\hline
12 & 6.26529 & 6.2690 & $+0.0037$ & $+0.0023$ & $-0.0080$ & $-0.0070$ & $-0.0169$\\
13 & 6.78765 & 6.7919 & $+0.0042$ & $+0.0030$ & $-0.0077$ & $-0.0068$ & $-0.0106$\\
14 & 7.31828 & 7.3256 & $+0.0073$ & $+0.0061$ & $-0.0075$ & $-0.0066$ & $-0.0126$\\
15 & 7.84978 & 7.8557 & $+0.0059$ & $+0.0048$ & $-0.0073$ & $-0.0066$ & $-0.0155$\\
16 & 8.37163 & 8.3797 & $+0.0081$ & $+0.0070$ & $-0.0072$ & $-0.0064$ & $-0.0087$\\
17 & 8.90563 & 8.9156 & $+0.0100$ & $+0.0090$ & $-0.0069$ & $-0.0064$ & $-0.0140$\\
18 & 9.43215 & 9.4406 & $+0.0085$ & $+0.0076$ & $-0.0067$ & $-0.0061$ & $-0.0119$\\
19 & 9.95659 & 9.9688 & $+0.0122$ & --- & --- & $-0.0061$ & $-0.0076$\\
\end{tabular}
\end{ruledtabular}
\end{table}

\textit{Errors of Eq.~(8).} (i) \emph{Drift.} Fits of $\Delta$Eq.~(8) to the sixteen deaths give
AIC $=-213.3$ for $a_0+a_1\kappa$ ($a_1=0.00163$, rms $0.0011$), $-188.3$ for $a_0+c/\kappa$ (rms $0.0025$), and $-218.2$
for $a_0+a_1\kappa+c/\kappa$ (rms $0.0009$): the error grows linearly in $\kappa$, so it is not a $1/\kappa$ term of the
WKB expansion. Its source is the $n=1$ pole. In the barrier-top solution, Eq.~\eqref{eq:ainSM}, the residues differ by the
derivative of $1/\Gamma$ and by $e^{ia(2\ln c+J)}e^{\pi a/2}$, which gives
\begin{equation}
\rho_1\equiv\frac{P_{1,H}}{P_{0,H}}\simeq\frac{\lambda_1}{\lambda_0}\,\frac{K(\lambda_1)}{K(\lambda_0)}\;i\,c^2e^{J},\qquad
c^2=\tfrac{8}{27}\sqrt{L_0}\ \Longrightarrow\ |\rho_1|\to\tfrac{4\sqrt3}{9}e^{J-2\pi}\kappa=0.0201\,\kappa,\quad\arg\rho_1\to\tfrac\pi2 .
\label{eq:rho1}
\end{equation}
Against the exact CAM ratio: $|\rho_1|=0.0478$/$0.0475$ ($\kappa=2$), $0.1084$/$0.1082$ ($5$), $0.1905$/$0.1904$ ($9.5$);
the phases agree to $0.10$, $0.04$, $0.02$~rad. The omitted term shifts the phase by $\arg(1+\rho_1)\approx0.02\kappa$ and
$\kappa_d$ by $\approx0.0017\kappa$. The same construction gives every pole: since $1/\Gamma(\frac12+ia)$ has derivative
$(-1)^nn!$ at $a=i(n+\frac12)$,
$\rho_n\equiv P_{n,H}/P_{0,H}\simeq(\lambda_n/\lambda_0)[K(\lambda_n)/K(\lambda_0)](ic^2e^J)^n/n!$, so
$|\rho_n|\approx|\rho_1|^n/n!$ to leading order, and $|\rho_2|\to\frac12|\rho_1|^2\propto\kappa^2$, $\arg\rho_2\to\pi$. Against
the exact CAM ratio, $|\rho_2|=0.00238$/$0.00239$, $0.00755$/$0.00755$, $0.01857$/$0.01857$ at $\kappa=3$, $6$, $9.5$ (phases
within $0.23$, $0.11$, $0.07$~rad). Near the horizon the higher poles therefore gain weight with frequency, although
$|K(\lambda_n)/K(\lambda_0)|\approx e^{-2\pi n}$; at $|\rho_1|\approx1$, i.e., $\kappa\approx50$, a one-pole description fails
and Eq.~\eqref{eq:exactdeath} needs many poles.
Adding it to Eq.~(8) (column (8$'$), no fitted parameter) leaves a constant
$-0.0064\pm0.0007$. With this term, the extrapolated error of Eq.~(8) itself is $+0.020$ at $\kappa=15$ and $+0.028$ at
$\kappa=20$ (linear fit: $+0.019$ and $+0.027$). (ii) \emph{Offset.} The on-axis criterion $\Phi_H^{(N)}$ converges by
$N=3$ to an offset of $-0.0083$ to $-0.0067$, while the off-axis criterion $Z^{(N)}$ converges by $N=3$ to the measured
deaths (Table~\ref{tab:offset}). At the zero of $Z^{(N)}$ the on-axis phase equals
$\arg[\sum P(0)/\sum P(\theta'_d)]-\arg[B(0)/B(\theta'_d)]$: the angular phase of the surface wave, $P_\nu(\cos\theta')$
with complex $\nu$, plus that of the background. The cell dies where the surface wave has decayed to the level of the
background, $|\sum P(0)/\sum P(\theta'_d)|=|\sum P/B|(0)$, i.e., at $\theta'_d=0.20$ to $0.07$. Test of the
extrapolation: on the axis at $\kappa=\kappa_d$, $\Phi^{(4)}(r)-\Phi_H^{(4)}=c\,t$ for $t=3\times10^{-2}$--$10^{-4}$, with
$c=10.1$, $17.2$, $26.0$ at $\kappa_d=3.08$, $5.20$, $7.85$; since $r_O-r_s\propto\kappa_d-\kappa$, the phase at the
core tends to $\Phi_H^{(4)}$, and the offset is not an $O(r-r_s)$ correction. It does not depend on the quadrature of the
background ($4\times10^{-9}$~rad when the Gauss rule, the panels and both cutoffs are refined). (iii) The WKB term
$c_1/\kappa$, $c_1=0.1026$ (from $\kappa\ge6$), shifts $\kappa_d$ by $-0.001$ to $-0.0005$. Items (i) and (ii) have
opposite signs and cancel near $\kappa\approx4$, which is why Eq.~(8) is most accurate for the first six deaths.

\begin{table}[ht]
\caption{On-axis phase criteria $\Phi_H^{(N)}=0$ and off-axis complex criteria $Z^{(N)}$ (Sec.~\ref{sec:phase}), minus the
measured death, for $N$ poles, and the angular phases at the zero of $Z^{(4)}$ (rad).}
\label{tab:offset}
\begin{ruledtabular}
\begin{tabular}{ccccccccc}
$\kappa_d$ & $\Phi^{(1)}$ & $\Phi^{(2)}$ & $\Phi^{(3,4,5)}$ & $Z^{(1)}$ & $Z^{(2)}$ & $Z^{(3,4,5)}$ & $\theta'_d$ & $\arg\frac{\sum P(0)}{\sum P(\theta'_d)}$; $-\arg\frac{B(0)}{B(\theta'_d)}$\\
\hline
2.01504 & $-0.0050$ & $-0.0084$ & $-0.0083$ & $+0.0046$ & $+0.0001$ & $+0.0002$ & 0.203 & 0.059; 0.041\\
3.08184 & $-0.0030$ & $-0.0089$ & $-0.0087$ & $+0.0065$ & $-0.0001$ & $0.0000$ & 0.158 & 0.059; 0.038\\
4.14023 & $-0.0014$ & $-0.0087$ & $-0.0085$ & $+0.0080$ & $+0.0002$ & $+0.0003$ & 0.136 & 0.066; 0.038\\
6.26529 & $+0.0023$ & $-0.0083$ & $-0.0080$ & $+0.0107$ & $-0.0001$ & $-0.0001$ & 0.101 & 0.062; 0.032\\
7.84978 & $+0.0048$ & $-0.0076$ & $-0.0073$ & $+0.0125$ & $0.0000$ & $0.0000$ & 0.086 & 0.062; 0.029\\
9.43215 & $+0.0076$ & $-0.0070$ & $-0.0067$ & $+0.0147$ & $+0.0002$ & $+0.0001$ & 0.074 & 0.060; 0.026\\
\end{tabular}
\end{ruledtabular}
\end{table}

On $[6,10]$ the measured spacings $0.5224$, $0.5306$, $0.5315$, $0.5219$, $0.5340$, $0.5265$, $0.5244$ compare with the
predicted $0.5229$, $0.5337$, $0.5301$, $0.5240$, $0.5359$, $0.5250$, $0.5282$; the eikonal spacing is $0.52866$. Over
all sixteen deaths the mean spacing is $0.52944\pm0.00002$, against $0.53028$ from Eq.~(8) (its roots $m=4$ and $19$).
For the first six deaths the mean spacing is $0.53239\pm0.00004$; the slope formula
$2\pi/(2\pi\,d\mathrm{Re}\lambda_0/d\kappa-D(r_s))$, with the derivative by central differences on a grid of step
$0.05$, ranges from $0.5301$ to $0.5362$ on $[2,4.7]$ (mean $0.5320$).

\textit{Births.} The same construction at a point $r<3/2$ on the axis gives the phase
$\Phi(\kappa,r)=2\pi\mathrm{Re}\lambda_0-\kappa D(r)+\beta(\kappa,r)$, with $\beta$ depending weakly on $r$ ($-0.05$
between $r=1.004$ and $1.4$), and the modulus $|P_0/B|(r)=|P_0/B|(r_s)F(r)$ with
\begin{equation}
F(r)=(1-b_c^2g)^{-1/4}\,\frac{1+\sqrt{1-b_c^2g}}{2r}\,e^{-j(r)/2},\qquad
j(r)=\ln\frac{2-\sqrt3}{2+\sqrt3}-\ln\frac{u-\sqrt3}{u+\sqrt3},\quad u=\sqrt{1+3/r},
\end{equation}
which agrees with the exact ratio to $0.05$--$0.6\%$ for $r\le1.4$. The model ``$\Phi=2\pi m$ and $|P_0/B|=1$'',
analytic except for $\mathrm{Re}\lambda_0(\kappa)$, reproduces the on-axis zeros (Table~\ref{tab:births}). It was not
run on $[6,10]$ before the scan, so it is a postdiction.

\begin{table}[ht]
\caption{Births: WKB model versus exact zeros $\kappa^*$, $z^*$ of $\Ep$ on the axis.}
\label{tab:births}
\begin{ruledtabular}
\begin{tabular}{cccccccc}
$m$ & 4 & 5 & 6 & 7 & 8 & 9 & 10\\
\hline
$\kappa_b$ (WKB) & 1.9554 & 2.4250 & 2.8926 & 3.3851 & 3.8590 & 4.3049 & 4.7403\\
$\kappa^*$ (exact) & 1.9470 & 2.4197 & 2.8829 & 3.3713 & 3.8549 & 4.3034 & 4.7382\\
$\Delta\kappa$ & $+0.008$ & $+0.005$ & $+0.010$ & $+0.014$ & $+0.004$ & $+0.002$ & $+0.002$\\
$\Delta z$ & $-0.005$ & $-0.002$ & $+0.001$ & $+0.004$ & $-0.002$ & $-0.001$ & $0.000$\\
\end{tabular}
\end{ruledtabular}
\end{table}

\textit{Meaning of $D(r)$.} $D(r)=\int_r^\infty\bigl(1-\sqrt{1-b_c^2f/x^2}\bigr)dx/f$ is the WKB phase deficit (per unit
$\kappa$) of the wave with critical angular momentum $L=b_c^2\kappa^2$, which passes over the barrier top, relative to
the incident wave $e^{-i\kappa x_*}$, accumulated between $r$ and infinity. It is real because the radicand has a double
zero at $x=3/2$. It is not the path of a ray: for $r<3/2$ the ray with $b=b_c$ does not exist (it winds onto the photon
sphere). The wave of the pole ($a=i/2$) differs from the critical wave by $O(1)$ in $\lambda$, which changes only the
modulus and the constant phase, not $D$.

\section{Regge poles, windows, and phase slopes}
\label{sec:tables}
Regge poles from Leaver's continued fraction ($2M=1$), checked against direct ODE integration: agreement
$6\times10^{-13}$ ($n=0$) to $2.5\times10^{-4}$ ($n=5$). Example at $\kappa=3$: $\lambda_0=7.832474+0.496948i$,
$\lambda_1=7.865654+1.487542i$, $\lambda_2=7.929898+2.468721i$. QNM check: $M\omega=0.248263-0.092488i$ ($l=1$),
$0.457596-0.095004i$ ($l=2$).

\begin{table}[ht]
\caption{Windows of existence of the first six cells. $\kappa_b$: birth (bisection of existence); $\kappa^*$: exact
zero of $\Ep$ on the axis; $z^*$: its position; $\kappa_d$: death (bisection).}
\label{tab:windows}
\begin{ruledtabular}
\begin{tabular}{ccccc}
$m$ & $\kappa_b$ & $\kappa^*$ & $z^*$ & $\kappa_d$\\
\hline
4 & $1.9453\pm0.0016$ & 1.94697 & $-1.1099$ & $2.01504\pm2\times10^{-4}$\\
5 & $2.4219\pm0.0031$ & 2.41966 & $-1.1877$ & $2.54590\pm2\times10^{-4}$\\
6 & $2.8844\pm0.0031$ & 2.88288 & $-1.2435$ & $3.08184\pm2\times10^{-4}$\\
7 & $3.375\pm0.025$   & 3.37126 & $-1.2529$ & $3.61621\pm2\times10^{-4}$\\
8 & $3.8625\pm0.012$  & 3.85490 & $-1.2714$ & $4.14023\pm4\times10^{-4}$\\
9 & $4.3125\pm0.012$  & 4.30340 & $-1.3174$ & $4.675\pm0.025$\\
\end{tabular}
\end{ruledtabular}
\end{table}

Table~\ref{tab:phase} gives the mean slope of the phase $\Phi=\arg P_0-\arg(-B)$ at fixed $r$ (linear fit over
$\kappa\in[1.5,5]$) and its decomposition into the kernel $K$, the residue ($-\arg A_\mathrm{in}'(\lambda_0)$), the radial
phase $\arg w(r;\lambda_0)$, the factor $\arg(2\lambda_0)$, and the background; the six columns add up to $d\Phi/d\kappa$. The kernel alone does not ``describe a full orbit'': only the sum of
all five terms is compared with $2\pi b_c-D(r)$. Near the horizon the sum of residue, radial phase, $\arg\lambda_0$ and background
reproduces $-D(r)$: at $r=1.11$, $-5.42+2.46-0.02-1.12=-4.10$ against $-D(1.11)=-4.07$. The kernel slope $16.25$ is
$2\pi\,d\mathrm{Re}\lambda_0/d\kappa$, slightly below $2\pi b_c=16.32$.

\begin{table}[ht]
\caption{Slope of the phase of $P_0$ relative to $-B$ and its decomposition (each entry is $d/d\kappa$ of the
corresponding phase).}
\label{tab:phase}
\begin{ruledtabular}
\begin{tabular}{cccccccc}
$r$ & $d\Phi/d\kappa$ & kernel $K$ & residue & radial phase & $\arg\lambda_0$ & background & $2\pi b_c-D(r)$\\
\hline
1.02 & 11.840 & 16.245 & $-5.423$ & $+3.959$ & $-0.021$ & $-2.920$ & 11.952\\
1.05 & 11.941 & 16.245 & $-5.423$ & $+3.116$ & $-0.021$ & $-1.976$ & 12.053\\
1.11 & 12.141 & 16.245 & $-5.423$ & $+2.462$ & $-0.021$ & $-1.121$ & 12.250\\
1.20 & 12.428 & 16.245 & $-5.423$ & $+2.071$ & $-0.021$ & $-0.443$ & 12.541\\
1.50 & 13.389 & 16.245 & $-5.423$ & $+1.774$ & $-0.021$ & $+0.815$ & 13.468\\
2.00 & 14.873 & 16.245 & $-5.423$ & $+2.038$ & $-0.021$ & $+2.035$ & 14.411\\
3.00 & 17.464 & 16.245 & $-5.423$ & $+2.982$ & $-0.021$ & $+3.682$ & 15.136\\
\end{tabular}
\end{ruledtabular}
\end{table}

\section{Controls: PEC, Mie sphere, dual sphere, Buchdahl star, membrane}
\label{sec:controls}
These comparisons separate what is due to duality from what is due to symmetry. In a lossless medium
$\nabla\cdot\avS=0$, so at a zero $\mathrm{tr}\,J=0$; generic zeros of 3D divergence-free flows are saddles and
saddle-foci~\cite{Tribelsky2022}, and a center requires $\lambda_\perp=0$. In dual media all zeros at linear
polarization are circles (O: exact centers, X: saddles) of index 0. The PEC sphere at $\kappa=3$ has 128 saddle-foci
$(1,2)$, 68 saddle-foci $(2,1)$, 97 saddles $(2,1)$, 37 saddles $(1,2)$ outside $r=1.05$. The degree of
$\avS/|\avS|$ on spheres $r=R_0$ ($R_0=1.05$, $1.5$, $2$, $3$, $4$, $5$) is $0$, $-1$, $-5$, $-1$, $+1$, $0$, integer to
$10^{-13}$, and in every shell between two spheres its change equals the sum of the indices of the zeros in that shell.
In the dual cases the degree is $0$ on every sphere, since circles of zeros have index $0$.

\begin{table}[ht]
\caption{Zeros of $\avS$ at linear polarization in the window $|x_i|\le5$, outside the object ($r\ge1.05$ for the PEC
sphere). For dual media all zeros are circles around the axis.}
\label{tab:zeros}
\begin{ruledtabular}
\begin{tabular}{lll}
object & zeros outside & structure\\
\hline
PEC, $\kappa=1$ & 33 points & axis 3, $E$ plane 6, $H$ plane 12, off the planes 12\\
PEC, $\kappa=2$ & 123 points & axis 3, $E$ plane 46, $H$ plane 46, off the planes 28\\
PEC, $\kappa=3$ & 330 points & axis 4, $E$ plane 124, $H$ plane 130, off the planes 72\\
Buchdahl, $\kappa=1$ & 4 circles (2 O, 2 X) & only behind the star ($z\ge1.85$)\\
Buchdahl, $\kappa=2$ & 16 circles (8 O, 8 X) & $z\in[-1.75,1.86]$\\
Buchdahl, $\kappa=3$ & 33 circles (16 O, 17 X) & $z\in[-1.75,3.94]$, $\rho\le2.13$\\
horizon, $\kappa=3$ & 1 pair (O, X) & $\rho=0.1425$ (O), $0.1315$ (X), $z\approx-1.081$\\
$\bar f$, $\epsilon=10^{-2}$, $\kappa=3$ & 1 pair & $\rho=0.0835$ (O), $0.0775$ (X), $z\approx-1.083$\\
$\bar f$, $\epsilon=10^{-3}$, $\kappa=3$ & 1 pair & $\rho=0.1379$ (O), $0.1273$ (X), $z\approx-1.082$\\
\end{tabular}
\end{ruledtabular}
\end{table}

\begin{figure}[ht]
  \includegraphics[width=0.6\textwidth]{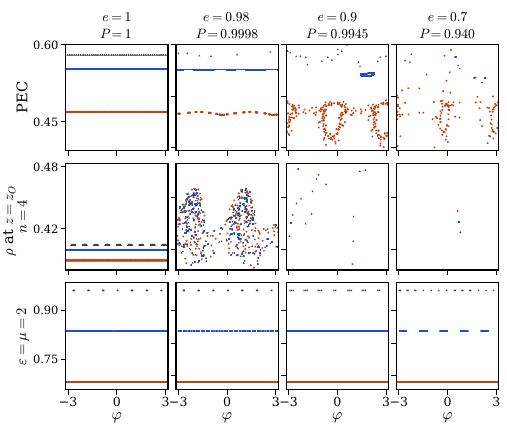}
  \caption{Poincaré sections of lines of the time-averaged $\avS$ at the same elliptical polarization
  $\hat{\mathbf x}+ie\hat{\mathbf y}$, $P=2e/(1+e^2)$, $e$ being the axis ratio of the ellipse. Each point is a
  crossing of a flow line with the plane $z=z_O$ through a vortex core ($\rho>\rho_O$); three starting points per
  panel. Rows: PEC sphere, dielectric sphere $n=4$, and dual sphere $\varepsilon=\mu=2$, all at $\kappa=3$; the column
  $e=1$ is circular polarization. Horizontal lines are invariant tori. Without duality the tori exist at $P=1$; as $P$
  decreases they are destroyed and resonant islands appear (we do not test for chaos). In the dual sphere they persist
  for every $P$. The cells differ in size ($|\mathrm O-\mathrm X|=0.39$ for the PEC cell, $0.04$ for the $n=4$ cell),
  so the thresholds of the two rows are not directly comparable. In the PEC and $n=4$ rows at smaller $P$, some flow lines leave the cell or their integration stops after a few crossings; these panels show short segments rather than full sections. Figure generated by AI-written scripts checked by the
  authors (see Use of AI tools).}
  \label{fig:poincare}
\end{figure}

\begin{itemize}
\item \textit{Circular polarization, non-dual objects.} A circularly polarized wave on the axis of any lossless
  axisymmetric scatterer gives an axisymmetric, integrable flow for that one polarization. PEC ($\kappa=3$): 20 O and
  16 X circles, no L cores; Mie $n=4$, $\kappa=3$: 7 O; weak scatterers ($n=1.5$, $\kappa\le3$; dual sphere at
  $\kappa=1.5$) have no recirculation cells. In one PEC cell ($q\in[0.093,2.29]$ at $P=1$) $q=3/2$ and $q=2/3$ both
  give $T(2,3)$, closing to $2.4\times10^{-10}$ and $3.0\times10^{-9}r_s$. Knots therefore are not specific to duality
  or gravity; what duality adds is the L core for every $P$.
\item \textit{Elliptical polarization} (Fig.~\ref{fig:poincare}): PEC tori survive at $P=0.9998$ near the core; at
  $P=0.9945$ they are destroyed and a resonant island appears; in the $n=4$ cell tori are destroyed already at
  $P=0.9998$. The symmetry-breaking terms are of order $|a_-|$, not $1-P$. The indicator is the residual of a smooth
  fit $\rho(\varphi)$ (Fourier modes $\le6$) compared with the scatter of the section points, not a Lyapunov exponent or a
  tolerance study: a torus is kept when the residual is below $\sim2\%$ of the scatter (e.g., $10^{-4}$ against
  $6\times10^{-3}$) and declared destroyed when it is comparable to it ($2\times10^{-2}$ against $2\times10^{-2}$--$7\times10^{-2}$).
\item \textit{Dual sphere} $\varepsilon=\mu=2$: Kerker condition $a_n=b_n$, no backscattering~\cite{ZambranaPuyalto2013};
  at linear polarization the $\cos2\varphi$ term of $\avS$ is $7\times10^{-18}$; one cell with an L core.
\item \textit{Buchdahl star}, $R=1.2$: cells near its surface are large ($0.1$--$0.5\,r_s$, circulating flux $10^{-2}$--$10^{-1}$),
  their cores are usually not L circles and their saddles are; in 88 cells (inside and outside the star) the types (core,
  saddle) are (non-L, L) 70, (L, non-L) 9, (L, L) 3, (non-L, non-L) 6.
\item \textit{Membrane} $\bar f$: the zero pair converges linearly, $\rho_O^\mathrm{hor}-\rho_O\approx4.6\,\epsilon$ for
  $\epsilon\le3\times10^{-3}$; loss per turn $\approx1.5\,\epsilon$. Since $\psi$ is not conserved there, this refers
  to the zeros of the meridional flow, not to tori. Regge poles of $\bar f$ at $\kappa=3$: $n=0$ converges linearly in
  $\epsilon$, $n=1$ converges, but $n=2$ does not in the range studied (distance $0.068$ at $\epsilon=10^{-3}$, $0.083$
  at $3\times10^{-4}$), possibly a manifestation of the spectral instability of higher
  overtones~\cite{JaramilloMacedoAlSheikh2021}.
\item \textit{Linear polarization, PEC.} A saddle-focus with eigenvalues $\lambda_\perp=0.00584$ (real part) and
  $\mathrm{Im}\lambda=10.94$ unwinds lines at $\gamma=0.0033$ per turn, against the linear value
  $2\pi\lambda_\perp/\mathrm{Im}\lambda=0.0034$; lines escape after $55$--$345\,r_s$.
\end{itemize}

\begin{figure}[ht]
  \includegraphics[width=\textwidth]{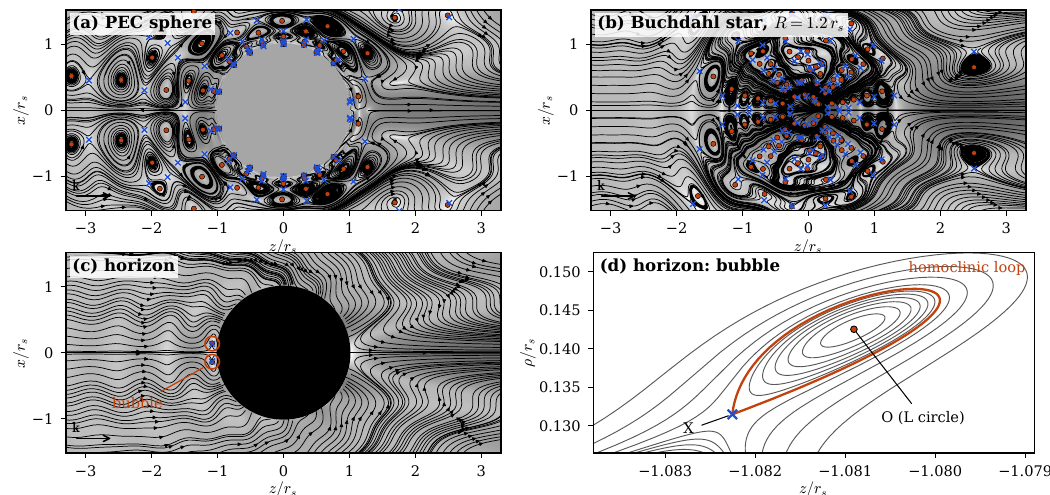}
  \caption{Lines of the time-averaged energy flow $\avS$ in the plane of $\mathbf{E}$ for a linearly polarized plane
  wave at $\kappa=3$. Gray scale: $\log_{10}|\avS|/S_\mathrm{in}$; filled circles: vortex centers (O); crosses: saddles
  (X). (a)~PEC sphere: isolated zeros, vortices are saddle-foci that unwind. (b)~Buchdahl star (surface at $r=1.2$):
  every zero is a circle around the axis cutting this plane twice. (c)~Horizon: one O--X pair of circles near the front
  pole. (d)~Close-up of the cell (coordinate axes, not to scale): contours of the Stokes function $\psi$; the
  separatrix through X is a homoclinic loop. These are lines of $\avS$, not of $\mathbf{E}$ or $\mathbf{H}$. Figure
  generated by AI-written scripts checked by the authors (see Use of AI tools).}
  \label{fig:three}
\end{figure}

\section{Canonical momentum}
\label{sec:canonical}
In optics the canonical (orbital) momentum density $\mathbf p_o\propto\mathrm{Im}[\mathbf E^*\cdot(\nabla)\mathbf E+\mathbf H^*\cdot(\nabla)\mathbf H]$
differs from $\avS$ by the spin term $\frac12\nabla\times\mathbf s$. In vacuum we recover
$\avS=\mathbf p_o+\frac12\nabla\times\mathbf s$ to $8\times10^{-11}$, but in the Plebański medium the identity fails in
both natural variants (with $\mathbf E,\mathbf H$ or with $\varepsilon\mathbf E,\mu\mathbf H$ in the first factors):
the relative residual is $1.5$--$3.8$ and $3.8$--$13$ at generic points near the front pole. The split into orbital and
spin parts is therefore ambiguous here. Its source is the anisotropy: for exact superpositions of six plane waves in
homogeneous dual media the identity holds up to a constant factor in an isotropic medium ($\varepsilon=\mu=2$:
$\mathbf p_o+\frac12\nabla\times\mathbf s=2\avS$ and $4\avS$ for the two variants, to $10^{-15}$), but in a uniaxial medium
($\varepsilon=\mu=\mathrm{diag}(10,10,1)$, the local ratio $\varepsilon_t/\varepsilon_r$ at $f=0.1$) no constant factor restores
it (residual after the best factor: median $0.52$ and $0.62$; for $\mathrm{diag}(2,2,1)$: $0.34$ and $0.31$). The
inhomogeneity adds a further term, since the vacuum derivation uses $\nabla\cdot\mathbf E=0$, whereas here
$\nabla\cdot(\varepsilon\mathbf E)=0$. Taken as an illustration, $\mathbf p_o$ has no O--X structure at the cell: in
$\rho\le0.45$, $-1.45\le z<-1.002$ the first variant has no zeros and the second one isolated O at $(0.067,-1.084)$,
$0.07\,r_s$ from the cell. Near the horizon $\mathbf p_o$ is dominated by the phase gradient of $t^{-i\kappa}$,
$|\mathbf p_o|\approx5$--$6$ at O and X, against $|\avS|\le0.03$. The cells are structures of the energy current
(Killing current), not of the canonical momentum.

\section{Checks of our own code}
\label{sec:checks}
All numbers refer to the horizon at $\kappa=3$, linear or circular polarization, with $l_\mathrm{max}=43=7\kappa+22$.
O lies at $(\rho,z)=(0.142537258,-1.080903534)$, X at $(0.131498119,-1.082255210)$;
$\psi_O-\psi_X=-2.4918982\times10^{-7}$; the circulating flux is $9.967593\times10^{-7}$ of the flux through
$\pi r_s^2$; $\psi$ is constant on the homoclinic loop to $1.6\times10^{-9}$ of $|\psi_O-\psi_X|$.

\textit{Convergence in $l_\mathrm{max}$.} Table~\ref{tab:conv} compares the cell computed with truncated partial-wave
sums against $l_\mathrm{max}=70$. For $l_\mathrm{max}=20$--$53$ the results are identical in floating-point arithmetic.
\begin{table}[ht]
\caption{Convergence of the cell at $\kappa=3$ relative to $l_\mathrm{max}=70$: shifts of O and X (in $r_s$) and
relative changes of $\psi_O-\psi_X$ and of the area $A$ enclosed by the homoclinic loop.}
\label{tab:conv}
\begin{ruledtabular}
\begin{tabular}{ccccc}
$l_\mathrm{max}$ & $|\Delta\mathrm O|$ & $|\Delta\mathrm X|$ & $\Delta\psi/\psi$ & $\Delta A/A$\\
\hline
10 & $2.4\times10^{-6}$ & $2.0\times10^{-6}$ & $-2.0\times10^{-4}$ & $-1.1\times10^{-4}$\\
12 & $1.2\times10^{-9}$ & $1.1\times10^{-9}$ & $-4.8\times10^{-7}$ & $-3.8\times10^{-7}$\\
15 & $3.1\times10^{-14}$ & $1.1\times10^{-13}$ & $-3.1\times10^{-11}$ & $+1.8\times10^{-9}$\\
20, 25, 30, 43, 53 & 0 & 0 & 0 & 0\\
\end{tabular}
\end{ruledtabular}
\end{table}

\textit{Three definitions of the Poynting vector.} The medium-picture vector $\avS$, the Killing-energy current
$J^i=-\sqrt{-g}\,T^i{}_t$ and the flux $S^{\hat\imath}$ measured by static observers in an orthonormal frame are related
by $\langle S\rangle^i=J^i/(r^2\sin\theta)$ for the coordinate components ($i=r,\theta,\varphi$; $J^i$ is a
coordinate density, and $r^2\sin\theta\,dr\,d\theta\,d\varphi$ is the flat volume element of the medium picture). The
physical spherical components of the medium-picture vector are $\avS_{\hat r}=\langle S\rangle^r$,
$\avS_{\hat\theta}=r\langle S\rangle^\theta$, $\avS_{\hat\varphi}=r\sin\theta\langle S\rangle^\varphi$, and
$\avS_{\hat r}=fS^{\hat r}$, $\avS_{\hat\theta,\hat\varphi}=\sqrt f\,S^{\hat\theta,\hat\varphi}$.
A field line of $S^{\hat\imath}$ in the spatial geometry has $dr/ds=\sqrt f\,S^{\hat r}$, $r\,d\theta/ds=S^{\hat\theta}$,
$r\sin\theta\,d\varphi/ds=S^{\hat\varphi}$; its coordinate velocity is $f^{-1/2}\avS$, so the two fields have the same
unparametrized integral curves and the same zeros. The static observers' fields diverge at $r_s$, while $J^i$ is
finite. Numerically, at 400 random points with $1.02<r<4$ the relations hold to $\le3.4\times10^{-16}$ relative to
$|\avS|$; O and X computed from $S^{\hat\imath}$ differ by $8.3\times10^{-16}$ and $2.2\times10^{-16}r_s$; the
homoclinic loop computed from $S^{\hat\imath}$ lies within a Hausdorff distance of $5.2\times10^{-6}r_s$ (the
integration tolerance). These checks test the consistency of the field transformation rather than the tetrad identity.

\textit{Conservation and normalization.} The energy flux through spheres, $\Phi(R)=\oint\avS_{\hat r}R^2d\Omega$, is
$-10.73984159$ for $R=1.01$, $1.0902$, $1.5$, $3$, $6$, $12$, with relative spread $2.9\times10^{-9}$ for linear and
circular polarization. The absorption cross section $\sigma_\mathrm{abs}=-\Phi/S_\mathrm{in}=21.47968\,r_s^2$ is $1.013$
times the geometric capture cross section $27\pi r_s^2/4$.

\textit{Maxwell residuals.} The fields were substituted directly into $\nabla\times\mathbf E=ik\mu\mathbf H$,
$\nabla\times\mathbf H=-ik\varepsilon\mathbf E$, $\nabla\cdot(\varepsilon\mathbf E)=\nabla\cdot(\mu\mathbf H)=0$ with
fourth-order central differences; residuals are relative to $k(|\mu\mathbf H|+|\varepsilon\mathbf E|)/2$. In the cell
(60 points, $h=10^{-3}$): curl $\le1.2\times10^{-6}$, divergence $\le4.1\times10^{-6}$. At O the curl residual falls as
$h^4$, from $3.8\times10^{-4}$ ($h=4\times10^{-3}$) to $5.6\times10^{-9}$ ($h=2.5\times10^{-4}$). On spheres $r=1.3$,
$2$, $4$: $\le3.6\times10^{-9}$; at $r=8$ a plateau of $1.1\times10^{-8}$, set by the cubic-spline interpolation of
the radial functions.

\section{Independent implementation}
\label{sec:codex}
\textit{Protocol.} A second implementation was written with a different AI system (OpenAI Codex CLI, model gpt-5.6-sol)
from three written task specifications. It did not open our code, notes, manuscript, reviews, figures, cache or
logs; the files it opened outside its own directory are listed in its record (\texttt{FILES\_READ.md}) and are only
our reference data for the comparison. It first derived and solved the problem, then froze its results
(\texttt{grid\_codex.csv}, \texttt{stagnation\_codex.csv}) by SHA-256 hashes (\texttt{FROZEN.md}, 2026-11-06,
20:02:43 CET), re-checked the hashes before opening the reference data, and only then compared. We re-computed both
hashes before writing this section and found them equal to the frozen values. One procedural deviation occurred before
the rules were known: the initial file-discovery command listed path names, including names in directories declared
forbidden later, without opening or searching any file content.

\textit{Method.} From the metric the code recovered symbolically the medium tensor and the Regge--Wheeler equation
for both polarizations. Radial functions start from a Frobenius series with only ingoing waves at the horizon, are
integrated with DOP853 (rtol $2\times10^{-10}$), and are matched at $r=250$ to tenth-order ingoing/outgoing Jost
expansions. The 3D field is summed to $l_\mathrm{max}=45$; $\psi$ is reconstructed by the trapezoidal rule with
$\psi=0$ on the front axis.

\textit{Internal tests of that code.} Flux balance $|A_\mathrm{in}|^2-|A_\mathrm{out}|^2-1$: $2.2\times10^{-9}$ ($l=1$,
$3$, $6$), $4.0\times10^{-7}$ ($l=10$); vacuum $\avS=(0,0,\frac12)$ to $2\times10^{-15}$; Maxwell residuals
$2.4\times10^{-10}$--$6.1\times10^{-10}$ with fourth-order convergence; flux through spheres $r=1.01$--$10$ constant to
$3.2\times10^{-10}$; fields converged to printed precision from $l_\mathrm{max}=20$.

\textit{Comparison at $\kappa=3$} (no change of axes, signs, amplitude or zero of $\psi$ was needed):
\begin{table}[ht]
\caption{Independent implementation versus our results at $\kappa=3$.}
\label{tab:codex}
\begin{ruledtabular}
\begin{tabular}{ll}
quantity & difference\\
\hline
position of O, of X & $7.5\times10^{-11}r_s$, $7.4\times10^{-11}r_s$\\
eigenvalue $\omega_O$, $\lambda_X$ (relative) & $1.2\times10^{-9}$, $2.8\times10^{-9}$\\
$\psi_O-\psi_X$ (relative) & $2.6\times10^{-8}$\\
separatrix (Hausdorff distance) & $1.1\times10^{-6}r_s$\\
$|\Delta\avS|/|\avS|$ on the $101\times101$ grid: median, 95th percentile, max & $8.4\times10^{-9}$, $2.8\times10^{-8}$, $1.0\times10^{-6}$\\
$|\Delta\psi|/|\psi_O-\psi_X|$ on the grid: median, max & $9.5\times10^{-4}$, $2.7\times10^{-3}$\\
$\sigma_\mathrm{abs}$ (relative) & $1.2\times10^{-9}$\\
linear polarization from $m=\pm1$: meridional part vs.\ circular, $|S_\varphi|$, $\varphi$ dependence & $2.8\times10^{-16}$, $2.8\times10^{-16}$, $3.8\times10^{-16}$\\
\end{tabular}
\end{ruledtabular}
\end{table}
The largest relative differences of $\avS$ sit near the zeros of the field, where the denominator is small. The larger
difference in $\psi$ reflects the reconstruction (trapezoidal rule versus our spline), and the Hausdorff distance
matches the resolution of the interpolated separatrix. The independent check covers the radial problem, the field,
and the cell at $\kappa=3$; it does not cover the knot construction or the CAM analysis.

\end{document}